\documentclass[useAMS,usenatbib]{mn2e}
\usepackage{graphicx}
\usepackage{times}
\usepackage{amssymb}
\usepackage{natbib}

\usepackage{pgf}
\usepackage{bm}

\def\gsim { \lower .75ex \hbox{$\sim$} \llap{\raise .27ex \hbox{$>$}} }
\def\lsim { \lower .75ex \hbox{$\sim$} \llap{\raise .27ex \hbox{$<$}} }

\begin{document}

\title[Origin of Galaxy Morphology]{The Origin of Disks and
  Spheroids in Simulated Galaxies}

\author[Sales et al.]{
\parbox[t]{\textwidth}{
Laura V. Sales$^{1}$,     
Julio F. Navarro$^{2}$,
Tom Theuns$^{3,4}$,
Joop Schaye$^{5}$,
Simon D. M. White$^{1}$,
Carlos S. Frenk$^{3}$,
Robert A. Crain$^{5}$ and
Claudio Dalla Vecchia$^{6}$
}
\\
\\
$^{1}$ Max Planck Institute for Astrophysics, Karl-Schwarzschild-Strasse 1, 85740 Garching, Germany\\
$^{2}$ Department of Physics and Astronomy, University of Victoria,
Victoria, BC V8P 5C2, Canada\\
$^{3}$ Institute for Computational Cosmology, Department of Physics,
University of Durham, South Road,  Durham, DH1 3LE, UK\\
$^{4}$ Department of Physics, University of Antwerp, Campus
Groenenborger, Groenenborgerlaan 171, B-2020 Antwerp, Belgium\\
$^{5}$ Leiden Observatory, Leiden University, PO Box 9513, 2300 RA
Leiden, Netherlands\\
$^{6}$ Max Planck Institute for Extraterrestrial Physics, Giessenbachstraße 1, 85748 Garching, Germany\\
}
\maketitle

\begin{abstract} The major morphological features of a galaxy are
  thought to be determined by the assembly history and net spin of its
  surrounding dark halo. In the simplest scenario, disk galaxies form
  predominantly in halos with high angular momentum and quiet recent
  assembly history, whereas spheroids are the slowly-rotating remnants
  of repeated merging events. We explore these assumptions using one
  hundred systems with halo masses similar to that of the Milky Way,
  identified in a series of cosmological gasdynamical simulations: the
  Galaxies - Intergalactic Medium Calculation ({\sc gimic}). At $z=0$,
  the simulated galaxies exhibit a wide variety of morphologies, from
  dispersion-dominated spheroids to pure disk galaxies. Surprisingly,
  these morphological features are very poorly correlated with their
  halo properties: disks form in halos with high and low net spin, and
  mergers play a negligible role in the formation of spheroids, whose
  stars form primarily in-situ. With hindsight, this weak correlation
  between halo and galaxy properties is unsurprising given that a
  minority of the available baryons ($\sim 40\%$) end up in
  galaxies. More important to morphology is the {\it coherent
    alignment of the angular momentum} of baryons that accrete over
  time to form a galaxy. Spheroids tend to form when the spin of
  newly-accreted gas is misaligned with that of the extant galaxy,
  leading to the episodic formation of stars with different kinematics
  that cancel out the net rotation of the system. Disks, on the other
  hand, form out of gas that flows in with similar angular momentum to
  that of earlier-accreted material. Gas accretion from a hot corona
  thus favours disk formation, whereas gas that flows ``cold'', often
  along separate, misaligned filaments, favours the formation of
  spheroids. In this scenario, many spheroids consist of the
  superposition of stellar components with distinct kinematics, age,
  and metallicity, an arrangement that might survive to the present
  day given the paucity of major mergers. Since angular momentum is
  acquired largely at turnaround, morphology depends on the early
  interplay between the tidal field and the shape of the material
  destined to form a galaxy.
\end{abstract}


\section{Introduction}
\label{SecIntro}

Galaxies exhibit a wide variety of morphologies, from spheroids
to disks to bars to peculiar galaxies of irregular shape. Many
physical properties, such as gas content, average stellar age, and the
rate of current star formation, are known to correlate with
morphology. Of such properties, the one that seems most tractable from
a theoretical perspective is the relative importance of organized
rotation in the structure of a galaxy. This is commonly referred to as
the disk-to-spheroid ratio, since stellar disks are predominantly
rotationally-flattened structures whereas spheroids have shapes
largely supported by velocity dispersion.

Since \citet{Hubble1926} published his original morphological
classification scheme, our understanding of the provenance of these
two defining features of galaxy morphology has been constantly
evolving. Spheroids were once thought to originate in the swift
transformation of an early-collapsing, non-rotating cloud of
gas  into stars \citep{ELS1962,Partridge1967,Larson1974}, whereas disks
were envisioned to result from the collapse of clouds with high
angular momentum and inefficient star formation
\citep{ELS1962,Larson1976}. The role of mergers as a
possible transformational mechanism was championed by
\citet{Toomre1977} and gained momentum as the hierarchical nature of
structure (and hence, galaxy) formation became accepted
\citep{White_Rees1978,Frenk1985}.

Further development of these ideas led to a broad consensus where
disks are thought to form at the center of dark matter
halos as a consequence of angular momentum conservation during the
dissipational collapse of gas \citep{Fall1980,Mo1998}, whereas
spheroids result predominantly from merger events \citep[see,
e.g.,][and references therein]{Cole2000}. Morphology is thus a transient
feature of the hierarchical formation of a galaxy: a disk galaxy may be
transformed into a spheroidal one after a major merger, but could then re-form a
disk through further gas accretion only to be later disrupted again by
another merger. Early galaxy formation simulations gave a
visually-compelling demonstration of this scenario, galvanizing
support for it \citep[see, e.g.,][]{SteinmetzNavarro2002}.

This consensus view has been broadly implemented in semi-analytic
models of galaxy formation, where the properties of galaxies are
deduced directly from the physical properties and assembly history of
their surrounding halos \citep[see,
e.g.,][]{Croton2006,Bower2006,Somerville2008}. For example, most
models assume that the specific angular momentum of galaxies and halos
are similar, and that the merger history of the halos dictates that of
the central galaxy.

Recent developments, however, have led to revisiting some of the
assumptions of the simple scenario outlined above. For example, it has
become clear that major mergers are rare, and therefore probably not
the primary formation mechanism of bulges and ellipticals. Instead,
``disk instabilities''
\citep{Efstathiou1982,Christodoulou1995,Mo1998}, as well as repeated
minor encounters, are now claimed to be the main formation path of
spheroids
\citep[e.g.,][]{Parry2009,Hopkins2010,DeLucia2011,Bournaud2011}. This
has helped to alleviate some tension between the observed evolution of
the early-type galaxy population and the major-merger rates predicted
by theory \citep{Bundy2007,Oesch2010}. However, questions might remain
open, as the estimation of merger times from observations is non-trivial
\citep{Lotz2011}.

Further scrutiny has come from direct simulation of hierarchical
galaxy formation. Conserving enough angular momentum during the
hierarchical assembly of a galaxy to form a realistic stellar disk has
been challenging
\citep[see, e.g.,][]{Navarro1995,NavarroSteinmetz1997}, as has been
pinning down the effect on morphology of repeated merging,
especially between gas-rich galaxies \citep[see,
e.g.,][]{Robertson2006,Governato2009}.

The inclusion of energetic feedback, needed to prevent the formation
of too many faint or overly massive galaxies, has added an extra level
of complexity to the problem, with a number of studies showing that
morphologies can be radically altered when even modest changes in the
strength of feedback or its implementation are introduced
\citep{Okamoto2005,Scannapieco2008,Ceverino2009,Sales2010,Agertz2011,Brook2011,Piontek2011}.
Moreover, the density threshold assumed for star
formation also changes the coupling between the stellar winds and the
surrounding gas, playing also a major role on the properties of
simulated galaxies\citep{Guedes2011}.

More recently, the {\it mode} of gas accretion has been recognized as
playing a potentially crucial role in galaxy morphology. Gas can flow
to galaxies largely unimpeded by shocks \citep{White1991} and may be
collimated by the filamentary structure of the cosmic web, especially
in low-mass systems and at high redshift
\citep{Keres2005,Dekel2006,vandeVoort2011}. This complex accretion
geometry has been hypothesized to promote the formation of disks by
feeding high angular momentum material directly to forming galaxies
\citep[see, e.g.,][]{Dekel2009,Brooks2009}.

Further theoretical progress demands increased sophistication in
numerical and semi-analytic modeling. From the simulation perspective,
most studies have focussed on individual systems picked 
according to what the authors believe would facilitate
the formation of a galaxy of predetermined morphology; for example, a
recent major merger to study ellipticals \citep[e.g.,][]{Meza2003} or
a quiet, rapidly-rotating halo to study spirals
\citep[e.g.,][]{Abadi2003a,Governato2007}. Note that this presupposes
the morphology of the resulting galaxy, and often results in the
tuning of star formation and feedback parameters until, unhelpfully,
results match prejudice.

Statistically-significant samples of galaxies selected in an unbiased
way and simulated at high resolution are needed for new insights, a
goal that, despite valiant efforts \citep{Croft2009, Sales2009c,
  Sales2010}, has so far proved beyond reach of even the fastest
computers and best algorithms.  The situation, however, is starting to
change, with the advent of simulations of volumes large enough to
include dozens of well-resolved $\sim L_\star$ galaxies
\citep{Crain2009,Schaye2010,Hahn2010,Cen2011,Vogelsberger2011}.

We explore these issues here using the {\sc gimic} gasdynamical
simulation series \citep{Crain2009}.  {\sc gimic} targeted several
carefully selected regions from the Millennium Simulation
\citep{Springel2005a} in an attempt to maximize the resolution of
individual galaxy systems while at the same time sampling a
cosmologically-significant volume. The first analyses of $z=0$  {\sc gimic}
galaxies \citep{Crain2010,Font2011b,McCarthy2011} show that they are
fairly realistic, so we feel confident that we can use them to gain
insight into the origin of galaxy morphology.

This paper is organized as follows. In Sec.~\ref{sec:numexp} we
present briefly the numerical method and simulations. We analyze the
morphologies of simulated galaxies and their origin in
Secs.~\ref{SecResults} and \ref{SecOrGxMorph}, respectively. We
summarize our main conclusions in Sec.~\ref{sec:concl}

\section{Numerical Simulations}
\label{sec:numexp}

The ``Galaxies-Intergalactic Medium Interaction Calculation'',
\citep[{\sc gimic};][]{Crain2009}, simulation series follows the
evolution of five nearly spherical regions of radius $\sim 20\, h^{-1}$ Mpc
each, selected from the {\it Millennium Simulation}
\citep{Springel2005a}. These regions were selected to sample
environments of different density, deviating by $(-2, -1, 0, +1,
+2)\sigma$ from the cosmic average, respectively, where $\sigma$ is
the {\it rms} mass fluctuation on $20 \, h^{-1} \rm Mpc$ scales. The
regions are spherical at $z=1.5$, and are simulated using the standard
zoom-in technique described in detail by, e.g., \citet{Power2003}.  We
provide here a basic summary of the main characteristics of the {\sc
  gimic} project, and refer the interested reader to \citet{Crain2009}
for a more comprehensive description.

{\sc gimic} uses a modified version of {\sc gadget-3}, a development of
the {\sc gadget-2} code \citep{Springel2005b} that includes new
modules to treat radiative cooling, star formation, chemical
enrichment, and energetic feedback. Radiative cooling is implemented
on an element-by-element basis and thus cooling rates evolve
self-consistently as a function of redshift, gas density, temperature
and chemical composition \citep{Wiersma2009a}. The runs also include a
uniform ionizing background \citep{Haardt2001}, with hydrogen- and
helium-reionization redshifts of $z=9$ and $z=3.5$, respectively.

Cold gas with densities exceeding $n_H =0.1$ cm$^{-3}$ becomes
eligible for star formation and is assumed to follow an effective
equation of state, $P \propto \rho^{4/3}$, in order to minimize
numerical artifacts in poorly-resolved regions
\citep{Schaye2008}. Stars are assumed to follow a Chabrier IMF
\citep{Chabrier2003}, and to form at a rate that depends on the local
gas pressure and that matches the Kennicutt-Schmidt law
\citep{Kennicutt1989,Kennicutt1998}.

Chemical enrichment is modeled as described by \citet{Wiersma2009b},
and tracks the synthesis of $11$ individual elements. As massive stars
explode as supernova (SN), they inject energy and metals into their
surroundings. This feedback is implemented, in practice, by using a
fraction $f_{SN}$ of the energy released by SN in order to modify the
velocity of a few ($\eta_w$) neighboring gas particles by introducing
a velocity ``kick'' of magnitude $V_w$ to each
\citep{DallaVecchia2008}.  These parameters are set in {\sc gimic} to
$f_{SN}=0.8$, $\eta_w=4$ and $V_w=600$ km/s, which results in a good
match to the peak of the global star formation rate density
\citep{Crain2009,Schaye2010}.

All {\sc gimic} runs adopt the same cosmological parameters as the
original {\it Millennium Simulation}, which were chosen to be
consistent with the WMAP-1 constraints: $\Omega_m=0.25$;
$\Omega_\Lambda=0.75$; $\Omega_b=0.045$; $n_s=1$; $\sigma_8=0.9$;
$H_0=100 \, h$ km s$^{-1}$ Mpc$^{-1}$; $h=0.73$.

The particle mass in the simulations is $1.4 \times 10^6 \, h^{-1} \,
M_\odot$ and $6.6 \times 10^6 \, h^{-1} \, M_\odot$ for the baryons
and dark matter, respectively. The gravitational softening is
initially fixed in comoving units, but is fixed at $z=3$ and thereafter to
$\epsilon=0.5 \, h^{-1}$ kpc (Plummer equivalent) in physical units.
We shall focus here on the two {\sc gimic} regions that have been run
to $z=0$ at this resolution: the $-2\sigma$ and $0\sigma$. As we
discuss below, aside from the expected difference in the number of
systems of given mass, we see no systematic dependence of our results
on the overdensity of the region, which may therefore be thought to
apply to average regions of the Universe.

We have used {\sc subfind} \citep{Springel2001a,Dolag2009} to identify
galaxies in the high-resolution regions of the {\sc gimic} runs. We
shall only consider in the analysis the {\it central} galaxies of
halos within a narrow range of virial\footnote{Virial quantities
  throughout this paper are defined at the radius enclosing 200 times
  the critical density for closure.} mass: $0.5<M_{200}/ 10^{12} \,
h^{-1} \, M_\odot<1.5$. This ensures homogeneity in the set of systems
selected for analysis and eliminates complications that may arise from
considering satellites of larger systems. At redshift $z=0$, these
criteria identify $38$ and $62$ galaxies in the $-2\sigma$ and
$0\sigma$ runs, respectively. Each of these halos is resolved with
roughly 200,000 particles (dark plus baryonic), allowing for a
reasonable estimate of the relative importance of the disk and
spheroidal components. None of the results we discuss here show
significant dependence on which region we consider, so we will group
the $100$ galaxies together without making distinction regarding
the {\sc gimic} run where they were identified.

\begin{center} \begin{figure} 
\includegraphics[width=84mm]{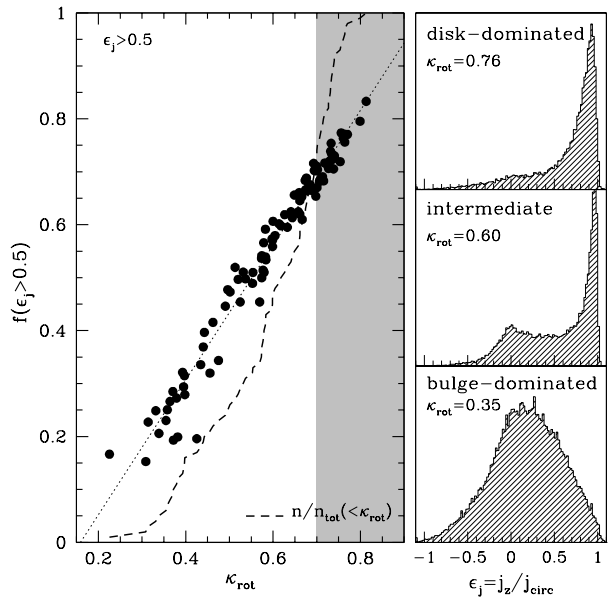} 
\caption{{\it Left:} The kinematic morphology parameter, $\kappa_{\rm
    rot}$, defined as the fraction of kinetic energy in organized
  rotation (eq.~\ref{EqKrot}), versus the fraction of stars with
  circularity parameter $\epsilon_j>0.5$. The cumulative fraction is
  shown with a dashed line. The shaded region ($\kappa_{\rm rot}>0.7$)
  indicates where ``disk dominated'' galaxies lie in this plot. {\it
    Right:} The distribution of circularities, $\epsilon_j=j_z/j_{\rm
    circ}(E)$, is shown for three galaxies with different values of
  $\kappa_{\rm rot}$.  }
\label{fig:kappa}
\end{figure}
\end{center}

\begin{center} \begin{figure*} 
\includegraphics[width=0.21\linewidth]{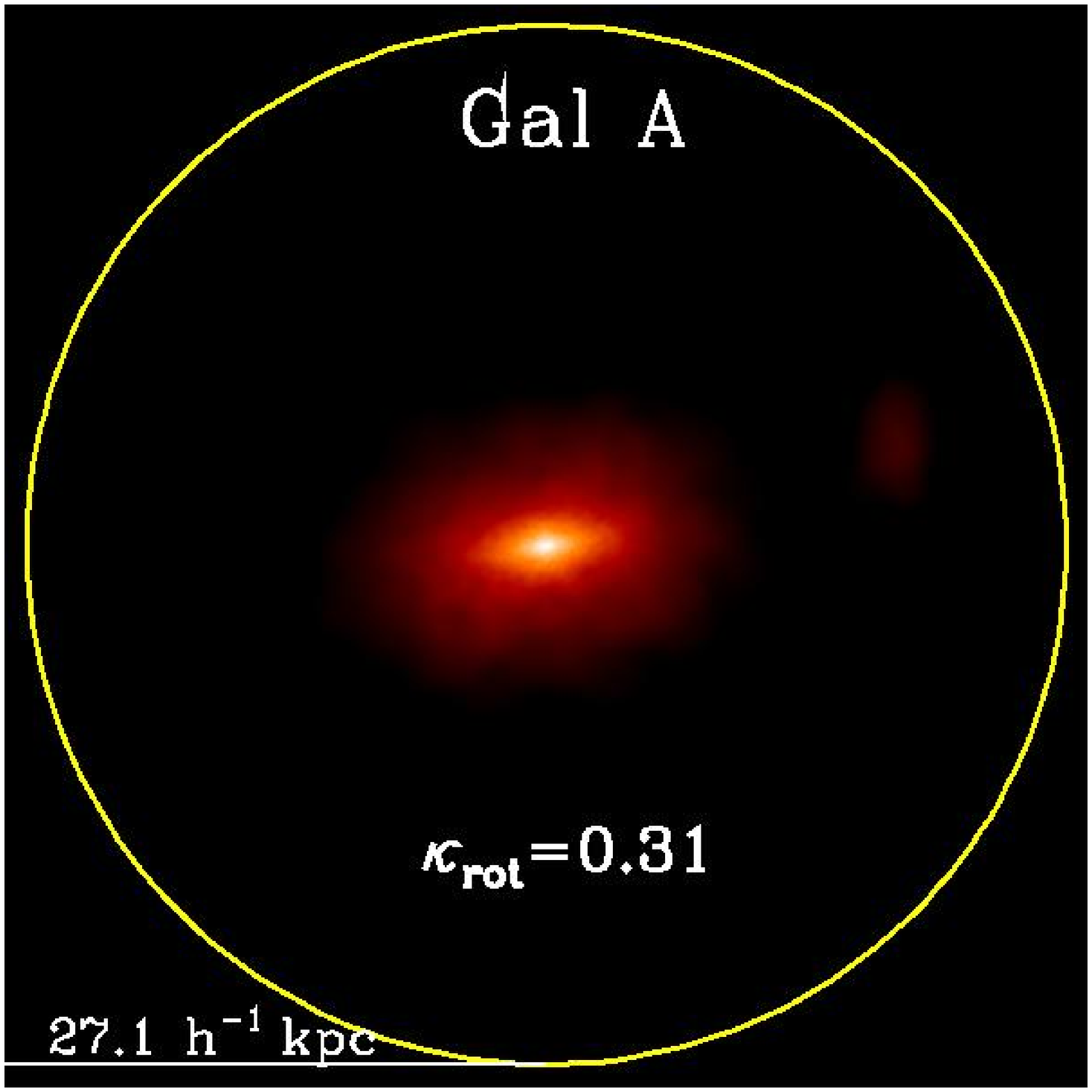} 
\includegraphics[width=0.21\linewidth]{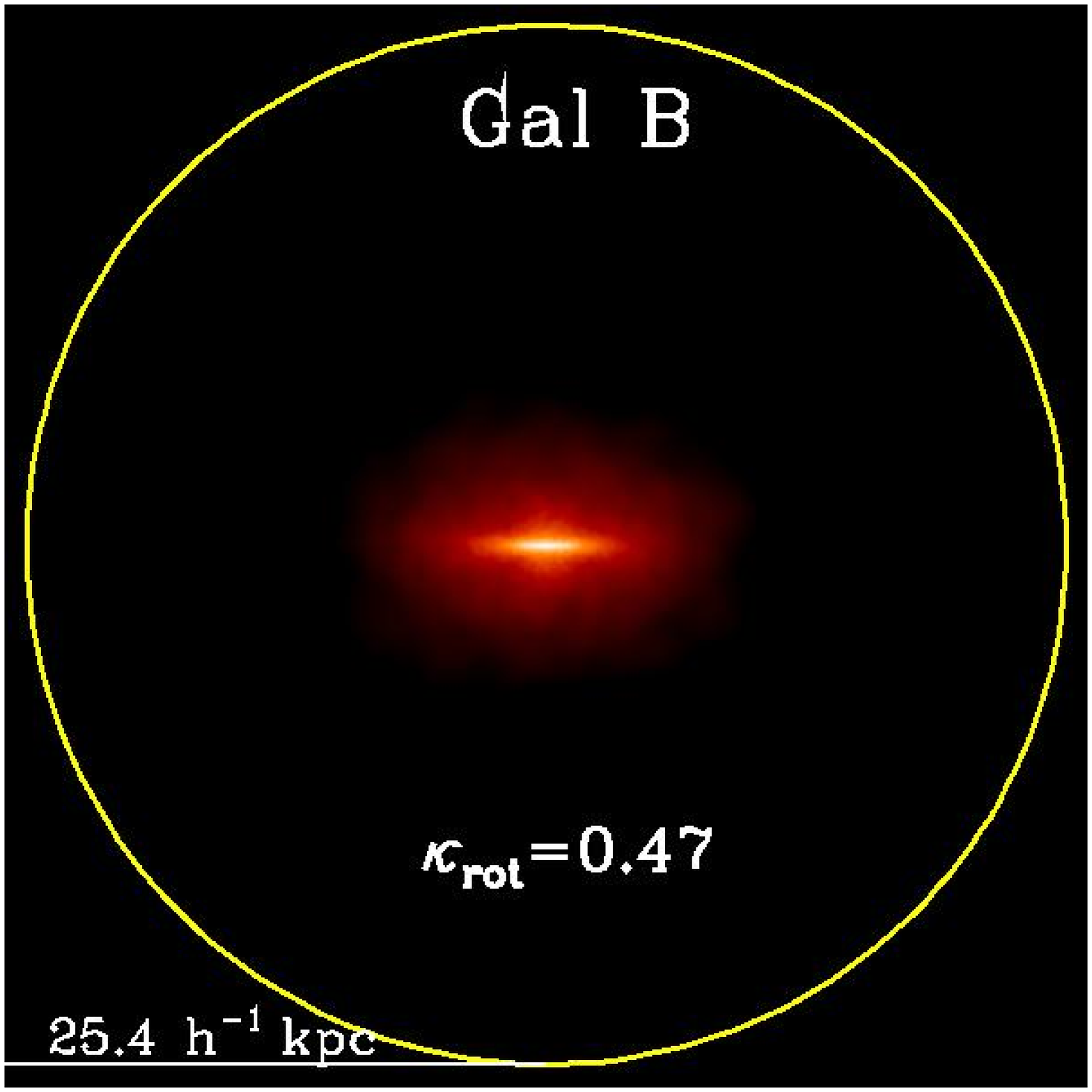} 
\includegraphics[width=0.21\linewidth]{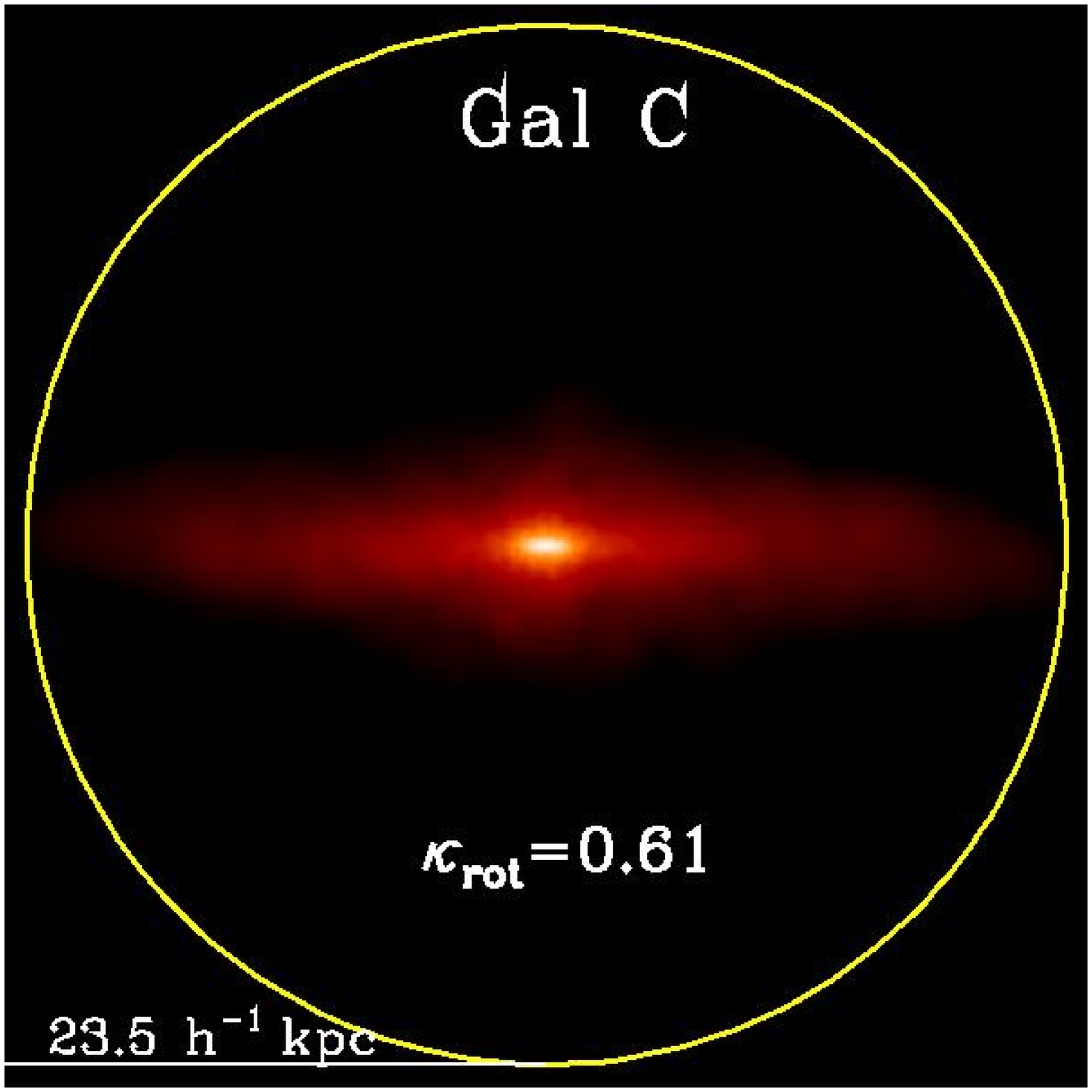} 
\includegraphics[width=0.21\linewidth]{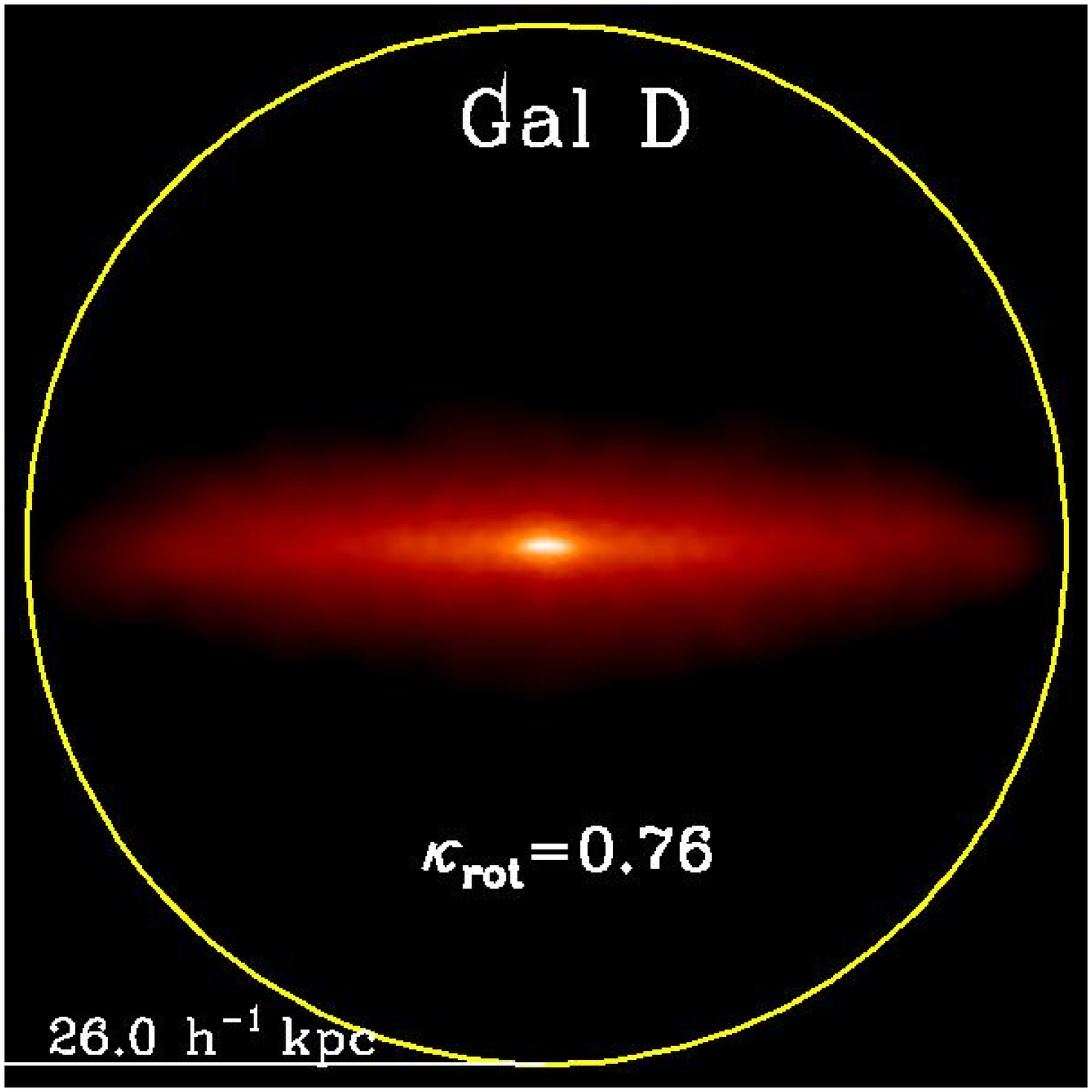}
\includegraphics[width=0.21\linewidth]{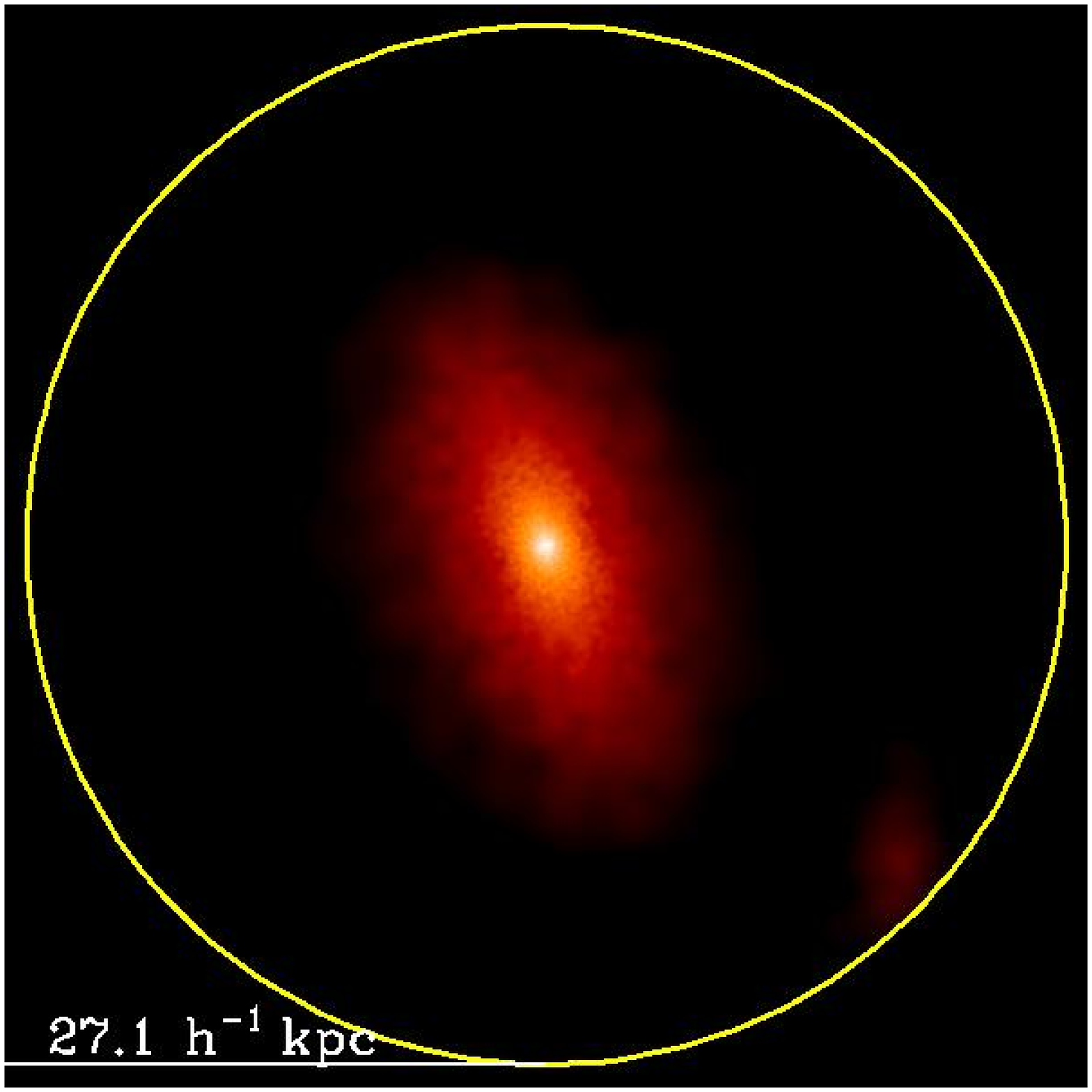} 
\includegraphics[width=0.21\linewidth]{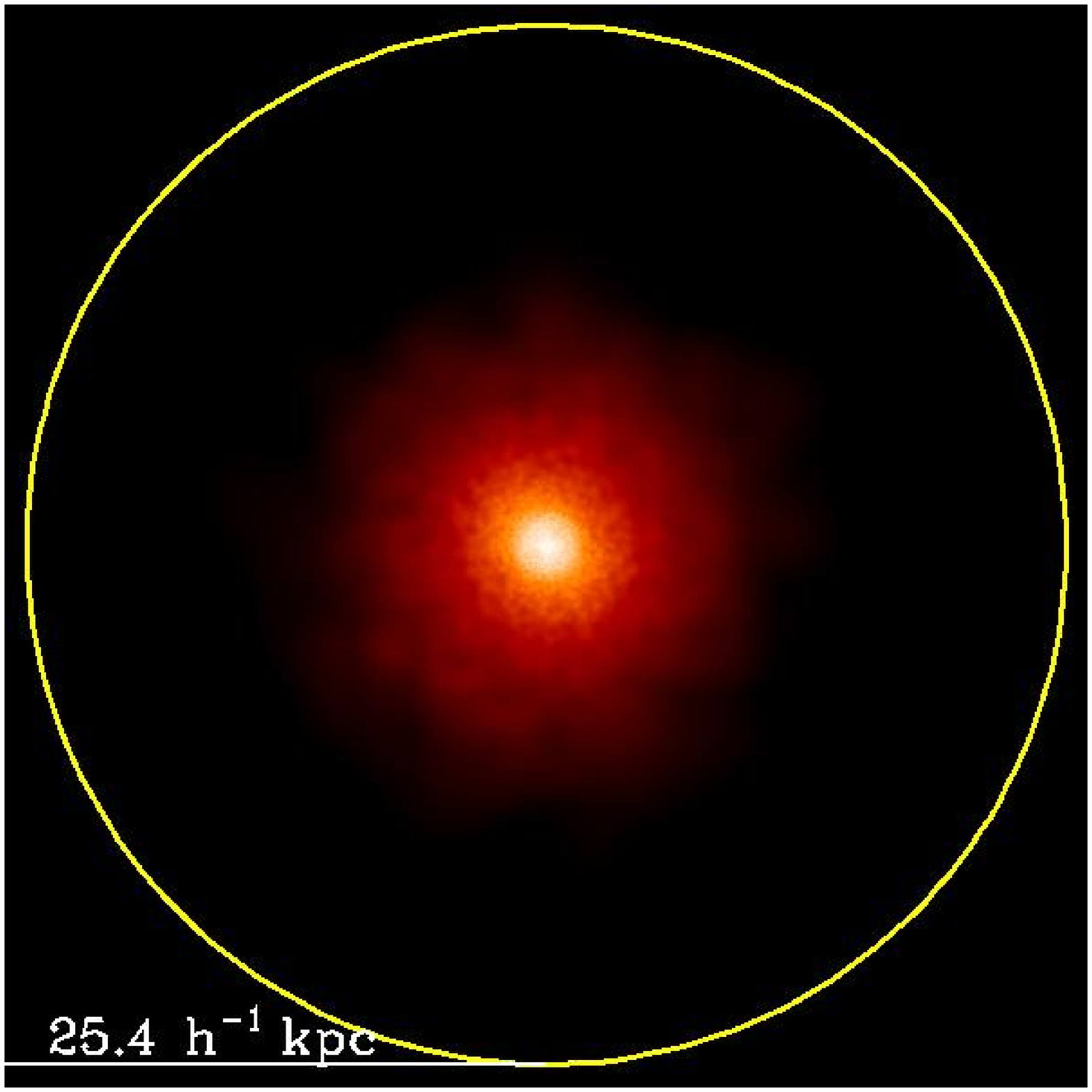}
\includegraphics[width=0.21\linewidth]{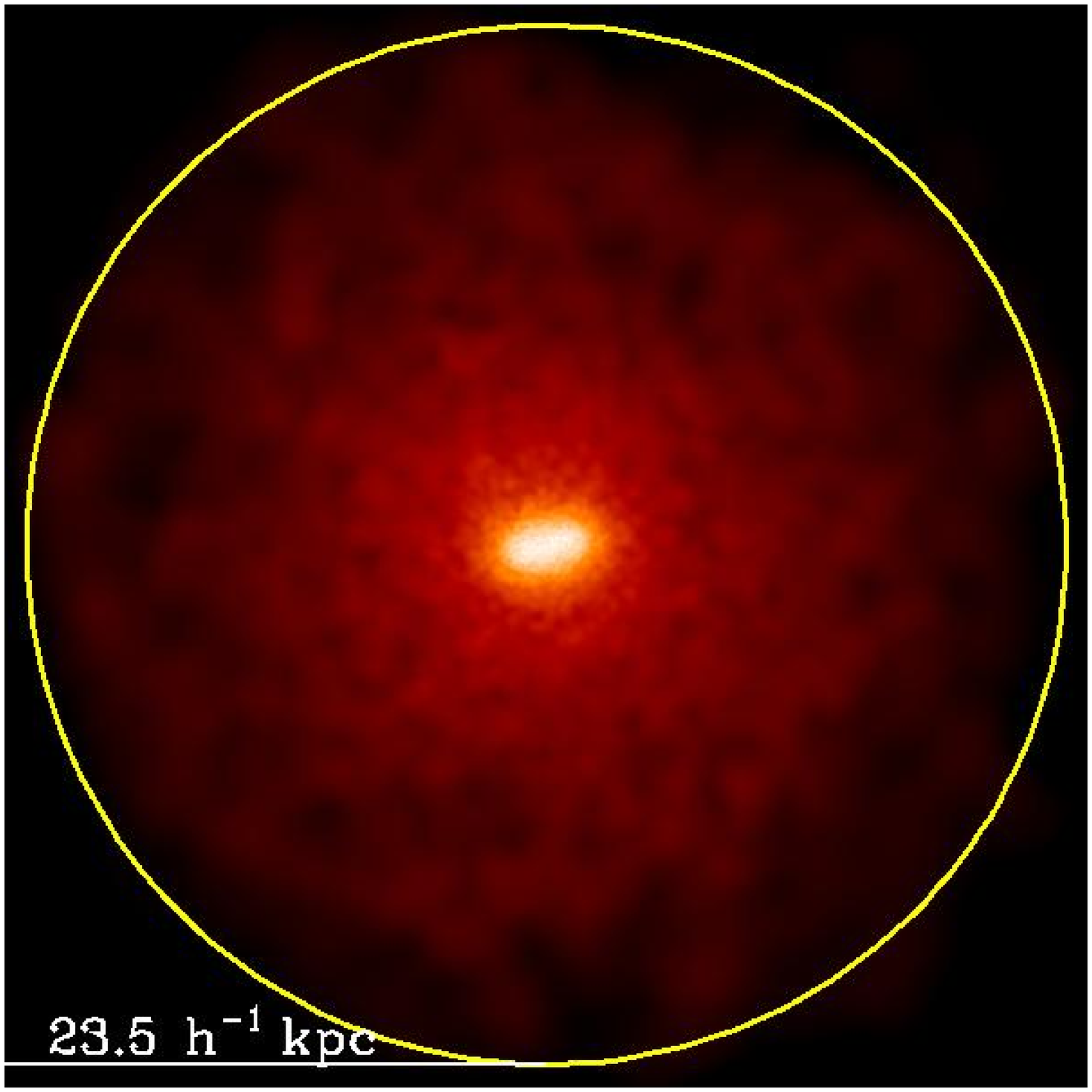} 
\includegraphics[width=0.21\linewidth]{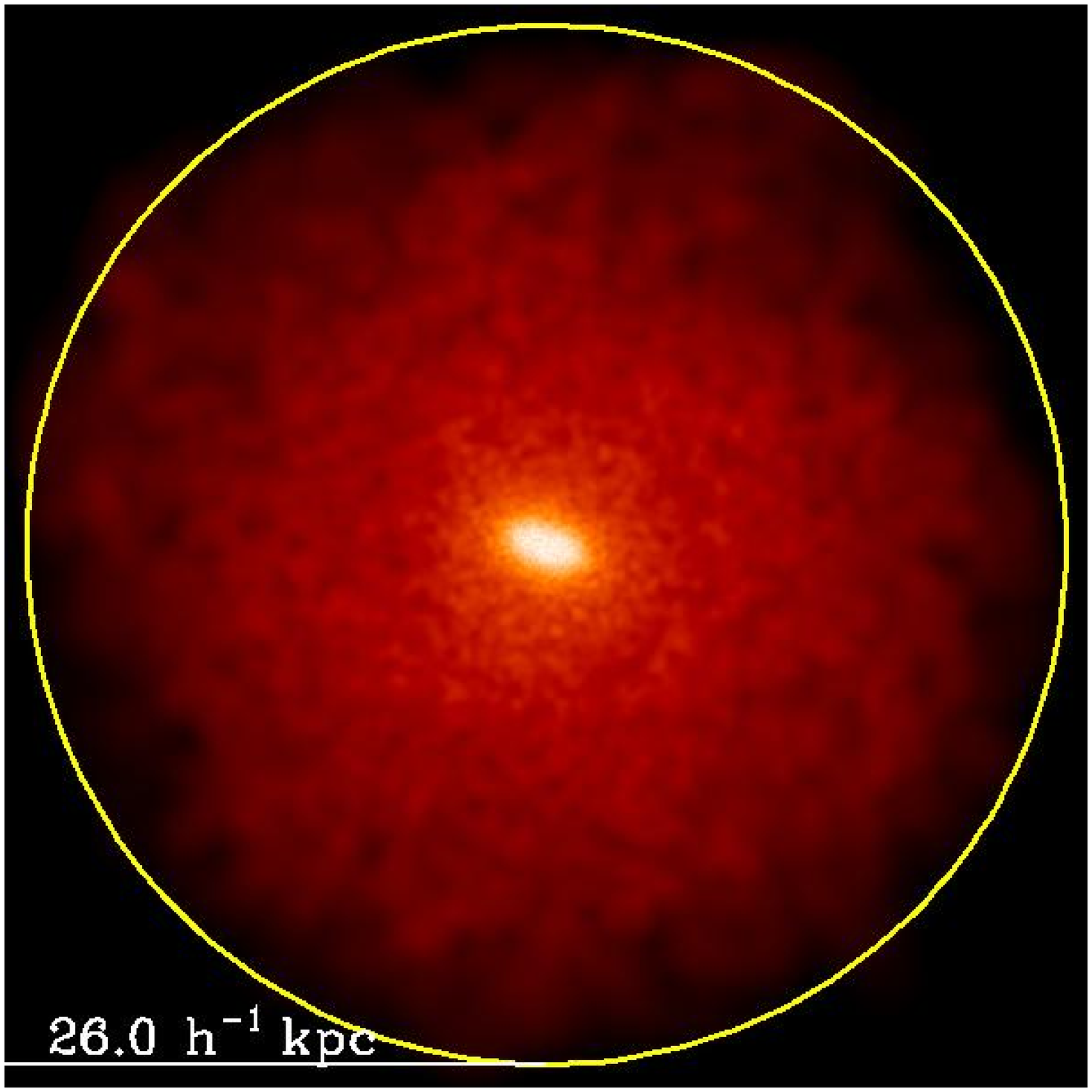}
\caption{Illustration of the structure of four galaxies in our sample
  with increasing degree of rotational support (left to right). The
  first and second rows show edge-on and face-on projections of the
  stellar distribution. The yellow circle marks the radius, $r_{\rm
    gal}=0.15 \, r_{200}$, used to define the galaxy.}
\label{FigImage}
\end{figure*}
\end{center}

\begin{center} \begin{figure*} 
\includegraphics[width=0.89\linewidth]{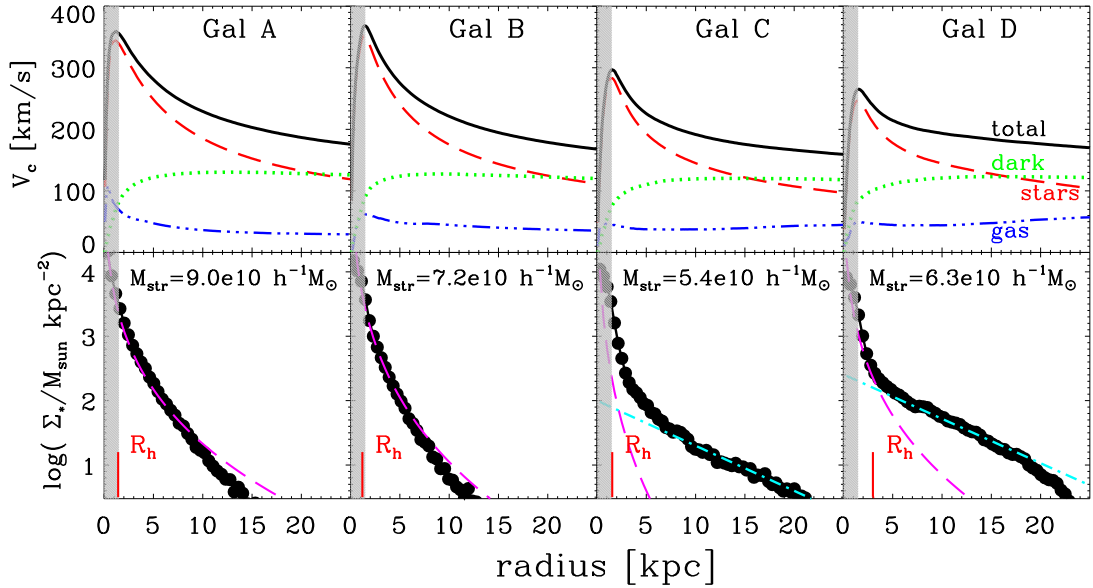}
\caption{Top and bottom rows show the circular velocity profile,
  $V_c(r)$, and the stellar surface density profiles, respectively, of
  galaxies A-D in Fig.~\ref{FigImage}. Red lines in the bottom
  panels indicate the stellar half-mass radius of each galaxy. Dashed
  magenta lines indicate de Vaucouleurs' $R^{1/4}$ profile
  fits. Straight dashed lines in blue indicate exponential profile
  fits. The shaded area highlights a radius equal to three
  Plummer-equivalent gravitational softening scalelengths. Note that
  spheroids can be well fit by a single $R^{1/4}$ law, but that
  disk-dominated galaxies show evidence of a de Vaucouleurs' spheroid
  plus an exponential disk. Disks are extended and have
 approximately flat circular velocity curves, spheroids tend to be denser and to
  have declining $V_c$ curves. }
\label{FigVcSigmaProf}
\end{figure*}
\end{center}

\begin{center} \begin{figure*} 
\includegraphics[height=180mm]{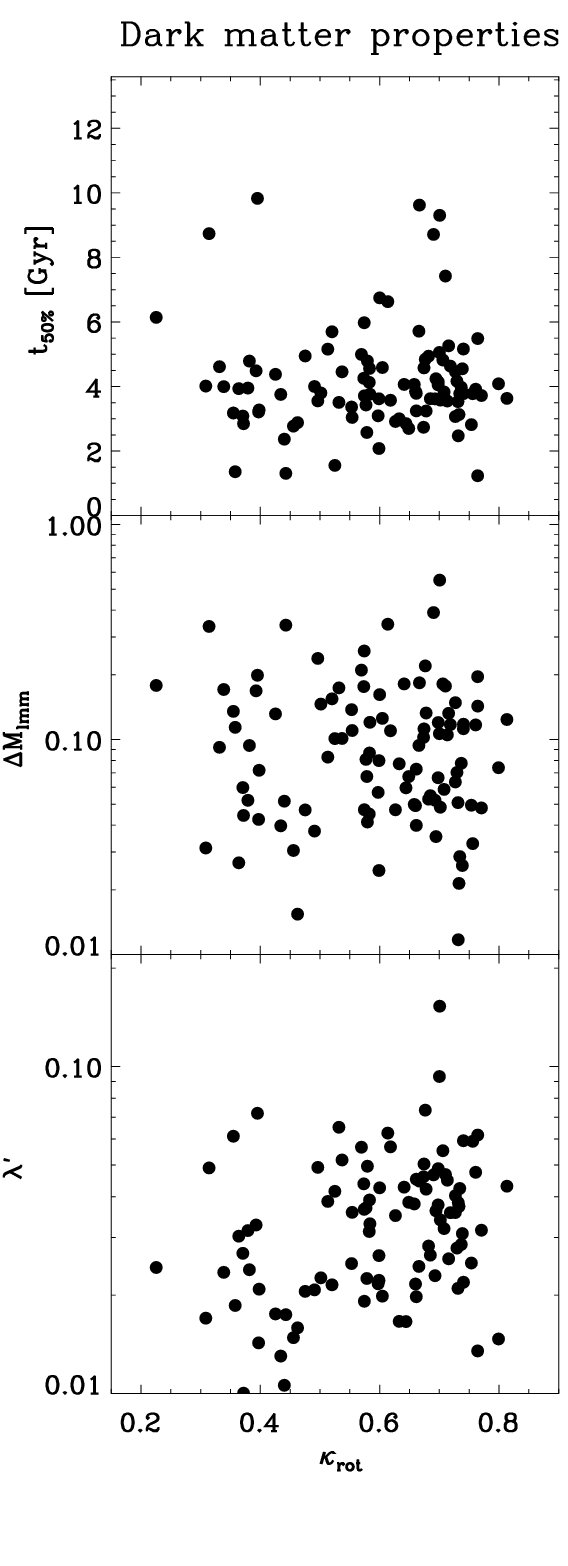}
\includegraphics[height=180mm]{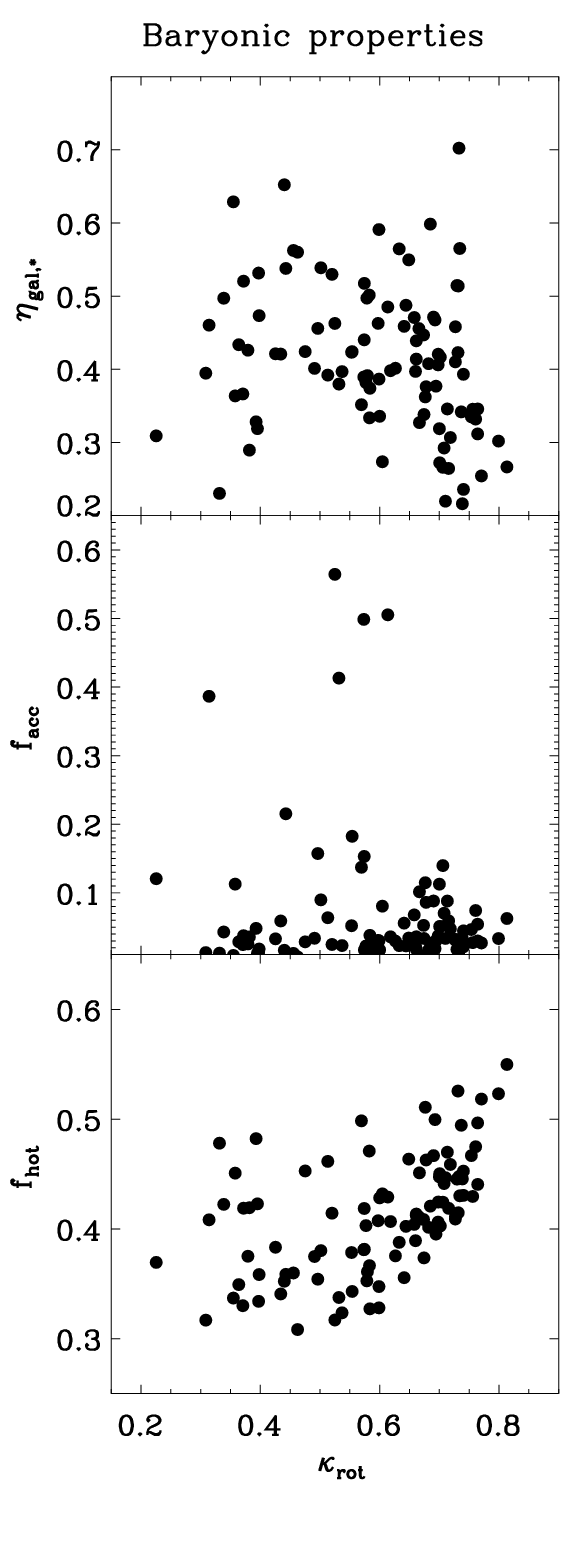} 
\caption{The kinematic morphology parameter, $\kappa_{\rm rot}$,
  versus a number of parameters characterizing the properties and
  assembly history of each galaxy and its surrounding halo. On the
  left, from top to bottom, $t_{50\%}$ is the half-mass halo formation
  time, in Gyrs; $\Delta M_{\rm lmm}$ is the maximum fraction of the
  final halo mass assembled in the single largest merger event after $z=3$;
  and $\lambda'$ is the dimensionless rotation parameter
  (eq.~\ref{EqLambdap}). On the right, the galaxy formation ``efficiency'', $\eta_{\rm
    gal,*}=M_{\rm gal}/(f_{\rm bar} M_{200})$; $f_{\rm acc}$ is the fraction
  of {\it accreted} stars (i.e., stars formed in galaxy progenitors
  other than the main one) and $f_{\rm hot}$ is the fraction of stars
  born out of gas that went through the ``hot phase'' (i.e.,
  $T_{\rm max}>10^{5.5}$ K). Correlation coefficients for each panel are given
  in Table~\ref{tab:correlation}.  See text for further details.}
\label{fig:correlations}
\end{figure*}
\end{center}

\section{Simulated Galaxy Morphologies}
\label{SecResults}

\subsection{Morphology estimates}
\label{SecMorphEst}

As discussed in Sec.~\ref{SecIntro}, we shall adopt a somewhat narrow
definition of morphology based on the importance of ordered rotation
in the structure of a galaxy. Although we refer to this as the ratio
of disk to spheroid, it should be noted that this may differ, at times
substantially, from traditional bulge-to-disk decompositions based on
photometric data. The latter are based on assumptions regarding
the shape of the brightness profile of disks, usually assumed to be
exponential, and spheroids, assumed to follow either de
Vaucouleurs or Sersic profiles. As discussed by \citet{Abadi2003a}
(see also \citealt{Scannapieco2010}), these assumptions are only
weakly fulfilled by simulated galaxies, and kinematic decompositions
can give rather different spheroid-to-disk ratios than photometric
ones. Photometric studies can also be affected by color gradients,
extinction, and projection effects \citep[see,
e.g.,][]{Governato2009}.  We avoid these complications by focusing our
analysis on kinematic data alone, although we plan to consider the
implications of our results for photometric studies in future work.

The importance of ordered rotation may be clearly appreciated from the
distribution of the stellar orbital {\it circularity} parameter,
$\epsilon_j=j_z/j_{\rm circ}(E)$, defined as the ratio of the
component of the specific angular momentum perpendicular to the disk
(i.e., aligned with the total angular momentum of the galaxy),
$j_z$, to that of a circular orbit with the same binding energy,
$j_{\rm circ}(E)$. Defined in this way, $\epsilon_j$ takes values in
the range ($-1$,$1$), where the extreme values correspond to the
counter- and co-rotating circular orbits in the symmetry plane of the
galaxy, respectively.

We show the $\epsilon_j$ distribution in the right-hand panels of
Fig.~\ref{fig:kappa} for three simulated galaxies, chosen to
illustrate three representative cases. The top panel corresponds to a
galaxy where most stars are in coplanar, nearly circular orbits, hence
the sharply-peaked distribution near $\epsilon_j=1$. The bottom
panel corresponds to a spheroidal galaxy where ordered rotation plays
little role; the $\epsilon_j$ distribution is broad and centered
around zero.  The middle panel corresponds to an intermediate case,
where a non-rotating bulge of stars is surrounded by a
well-defined thin disk. A simple quantitative measure of morphology
can therefore be constructed by the fraction of stars with circularities
exceeding a fixed fiducial value, such as $f(\epsilon_j>0.5)$.

Although conceptually simple, $\epsilon_j$ distributions are not easy
to relate to observation, so a simpler quantitative measure of
morphology is desirable. One alternative is the fraction of kinetic
energy invested in ordered rotation,
\begin{equation}
 {\kappa_{\rm rot}={K_{\rm rot} \over K}={1 \over K} \sum {1 \over 2}
   m \bigl({j_z\over R}\bigr)^2}.
\label{EqKrot}
\end{equation}
$\kappa_{\rm rot}\sim 1$ for disks with perfect circular motions, and
is $\ll 1$ for non-rotating systems. As Fig.~\ref{fig:kappa} makes
clear, $\kappa_{\rm rot}$ correlates extremely well with the fraction
of stars with $\epsilon>0.5$.  In what follows, we shall use
$\kappa_{\rm rot}$ to rank galaxies according to the importance of
their rotationally-supported components.

For convenience, we shall hereafter refer to galaxies with
$\kappa_{\rm rot}<0.5$ and $\kappa_{\rm rot}>0.7$ as spheroid- or
disk-dominated, respectively. The first group makes up $\sim 25\%$ of
the sample; the second group comprises another $\sim 30\%$. The
remainder consist of intermediate types where both rotation and
velocity dispersion play a comparable structural role. It is important
to note that our sample contains galaxies spanning a wide range in
$\kappa_{\rm rot}$, from ``pure disk'' systems with a negligible
fraction of stars in counter-rotating orbits (i.e., $j_z<0$) to
spheroids with little trace of rotational support.

\subsection{Examples of galaxy morphologies}
\label{SecGxMorph}

Fig.~\ref{FigImage} shows four examples chosen to illustrate the
structure of galaxies with various values of $\kappa_{\rm rot}$. The
panels show edge-on and face-on projections of each galaxy, colored by
stellar surface mass density on a logarithmic
scale. Fig.~\ref{FigVcSigmaProf} shows circular velocity profiles
and (face-on) stellar surface density profiles. The degree of
rotational support increases from left to right: the leftmost and
rightmost are spheroid- and disk-dominated systems, respectively,
while the two middle ones are intermediate-type objects. Labels in
each panel indicate, for each galaxy, the stellar mass within the
radius, $r_{\rm gal}=0.15 \, r_{200}$, used to define the central
galaxy.  Table~\ref{tab:table1} lists some physical parameters of
galaxies A-D.

Figs.~\ref{FigImage} and \ref{FigVcSigmaProf}, together with
Table~\ref{tab:table1}, show that simulated galaxies have several
properties in common with nearby ellipticals and
spirals. Spheroid-dominated galaxies are gas-poor, dense stellar
systems with declining circular velocity curves, whereas
disk-dominated galaxies are richer in gas, more spatially extended,
and have nearly-flat circular velocity curves. Interestingly, from the
point of view of surface density profiles, spheroids are
single-component systems well approximated by de Vaucouleurs'
$R^{1/4}$ law (dashed magenta lines in the bottom panels of
Fig.~\ref{FigVcSigmaProf}).  Disk-dominated systems, on the other hand,
have more complex profiles, with a central $R^{1/4}$ spheroid
surrounded by an exponential component that increases in importance in
step with $\kappa_{\rm rot}$. Like most spirals, they are well
approximated by the sum of a de Vaucouleurs' spheroid (dashed magenta
lines) and an exponential law (dashed cyan lines).

As discussed in detail by McCarthy et al (in preparation), these
similarities with observation actually extend to quantitative
comparisons with observed scaling laws, such as the Tully-Fisher
relation or the Fundamental Plane.  The agreement between simulated
galaxies and observation is encouraging, and suggests that the origin
of the morphological diversity of simulated galaxies can provide
insight into what determines the relative importance of disks and
spheroids in real galaxies.

A few caveats should also be mentioned. A wide variety of
morphological types does not automatically guarantee reproducing the
right morphological mix of galaxies in this mass range. Indeed, a
casual inspection suggests that spheroids might be over-represented in
our sample. However, because most morphological classifications are
based on photometric data, exploring this issue in detail would
require synthesizing ``observations'' of the simulated galaxies in
various bands and analyzing them in the same way as observed samples,
which is beyond the scope of this work. Moreover, as discussed in
\citet{Crain2009}, the {\sc gimic} galaxy stellar mass function
differs substantially from the observed one, implying that such
exercise would be inconclusive, regardless of its results. Our main
goal is thus not to test the viability of the particular star
formation/feedback implementation adopted in {\sc gimic}, but rather
to learn about the various mechanisms responsible for the origin and
relative importance of rotationally-supported versus
dispersion-supported stellar components in a galaxy.

\begin{center} \begin{figure}
 \includegraphics[width=84mm]{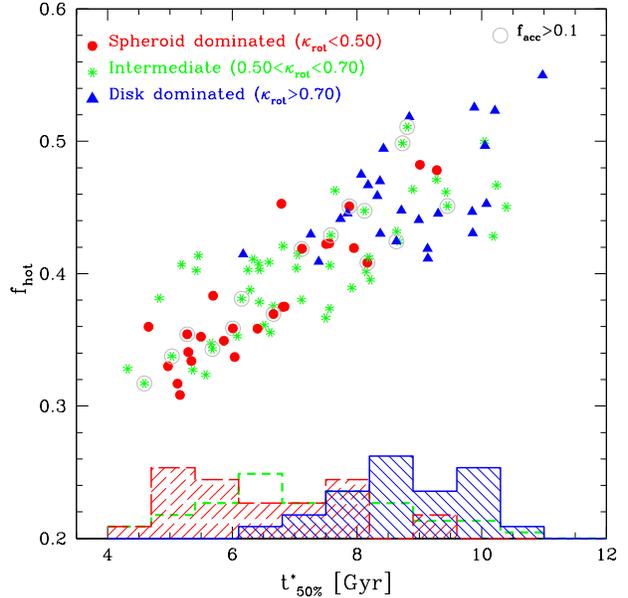} 
 \caption{The fraction of stars born from gas that went through the
   ``hot phase'' (i.e., $T_{\rm max}>10^{5.5}$ K) versus the median
   formation time of stars in the galaxy, $t^*_{50\%,}$. Symbols
   encircled in grey correspond to systems where more than $10\%$ of
   stars were accreted in merger events.  Each galaxy is color coded
   according to its morphology: red dots, green asterisks and blue
   triangles correspond to spheroid-dominated, intermediate, and
   disk-dominated systems, respectively, as labeled by the
   $\kappa_{\rm rot}$ parameter.  Histograms show the distribution of
   median formation time for each of these three groups.  The tight
   correlation between $f_{\rm hot}$ and $t^*_{50\%}$ suggest that gas
   able to reach the ``hot phase'' take longer to accrete and to be
   transformed into stars that gas accreted ``cold''. Late gas accretion
   clearly favors the assembly of stellar disks.}
\label{figs:age_fhot}
\end{figure}
\end{center}

\begin{center} \begin{figure} 
\includegraphics[width=84mm]{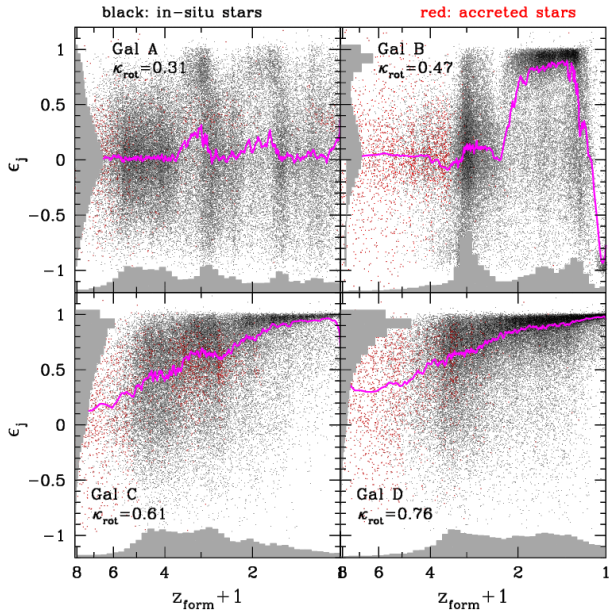} 
\caption{Formation time of stars (expressed in terms of redshift)
  versus the circularity parameter, $\epsilon_j$, measured at $z=0$,
  for the four galaxies illustrated in Fig.~\ref{FigImage}. Stars
  formed in the main progenitor (in situ) are shown in black, accreted
  stars in red. The magenta curve tracks the median
  circularity as a function of formation time. The two
  spheroid-dominated galaxies show signs of episodic star formation
  events that lead to the presence, at $z=0$, of stellar populations
  with distinct angular momentum properties. Disks, on the other hand,
  tend to form more gradually over time and to be dominated by a
  single population with coherently-aligned angular momentum.}
\label{fig:tform}
\end{figure}
\end{center}

\section{The origin of simulated galaxy morphologies}
\label{SecOrGxMorph}

\subsection{Dependence on dark halo properties}
\label{SecDepHalo}

As discussed in Sec.~\ref{SecIntro}, stellar disks are 
expected to form at the centers of halos with quiet recent accretion
histories and high angular momentum. Halos that have been relatively
undisturbed by recent major mergers tend to form earlier, so we may
also expect stellar disks to inhabit halos with early formation times.

We explore this in Fig.~\ref{fig:correlations}, where the left panels
show the dependence of $\kappa_{\rm rot}$ on (i) the halo formation
time, $t_{50\%}$ (when the most massive halo progenitor reaches half
the final halo mass); on (ii) the fraction of halo mass accreted in
the single largest merger after $z=3$, $\Delta M_{\rm lmm}$, and on
(iii) the dimensionless rotation parameter,
\begin{equation}
\lambda'={1 \over \sqrt{2}} {J \over M_{200} V_{200} r_{200}}, 
\label{EqLambdap}
\end{equation} 
where $J$ is the total angular momentum of the halo
\citep{Bullock2001b}.

None of these parameters correlates strongly with galaxy morphology
(see Table~\ref{tab:correlation}). Disks form in halos with low and
high spin parameter; in halos that collapse early and late, and even
in halos that have accreted a substantial amount of mass in merger
events. The same applies to spheroids, except perhaps for a weak
tendency to prefer halos with slightly lower-than-average $\lambda'$.

Fig.~\ref{fig:correlations} also shows that major mergers are uncommon
during the formation of halos in the narrow mass range
considered here; $0.5<M_{200}/10^{12} \, h^{-1}\, M_\odot<1.5$.  Most
systems have accreted less than $20\%$ of their final mass in a single
merger since $z=3$, and these events seem unrelated to the morphology
of the central galaxy at $z=0$. 

Finally, morphology also seems unrelated to the fraction of baryons
within the virial radius that collects to form the galaxy. This is
illustrated in the top-right panel of Fig.~\ref{fig:correlations},
where we plot $\kappa_{\rm rot}$ vs the galaxy formation
``efficiency'' parameter, $\eta_{\rm gal,*}=M_{gal,*}/(f_{\rm bar}\,
M_{200})$, where $f_{\rm bar}=\Omega_b/\Omega_m=0.175$ is the universal
baryon fraction. Although we consider halos in a narrow mass range,
the efficiency of galaxy formation varies from $20\%$ to $70\%$ (with
an average of $\langle \eta_{\rm gal,*} \rangle = 40\%$) and appears
to have little influence on the morphology of the central galaxy
\footnote{ We note that abundance matching models \citep[see,
  e.g.,][]{Moster2010,Guo2010} demand an even lower galaxy formation
  efficiency to match the galaxy stellar mass function, which would
  weaken even further the link between the properties of a central
  galaxy and that of its surrounding halo.}
although there is a weak tendency for disky objects to prefer lower
values of $\eta_{\rm gal,*}$.

Thus, contrary to simple expectations, spheroids {\it can} form in
quiescent halos, and disks {\it can} form in halos of scant angular
momentum content. Simple predictions of the morphology of a galaxy
based on the properties and assembly history of its surrounding dark
halo will thus often be wrong.

\begin{center} \begin{figure*} 
 \includegraphics[width=0.475\linewidth]{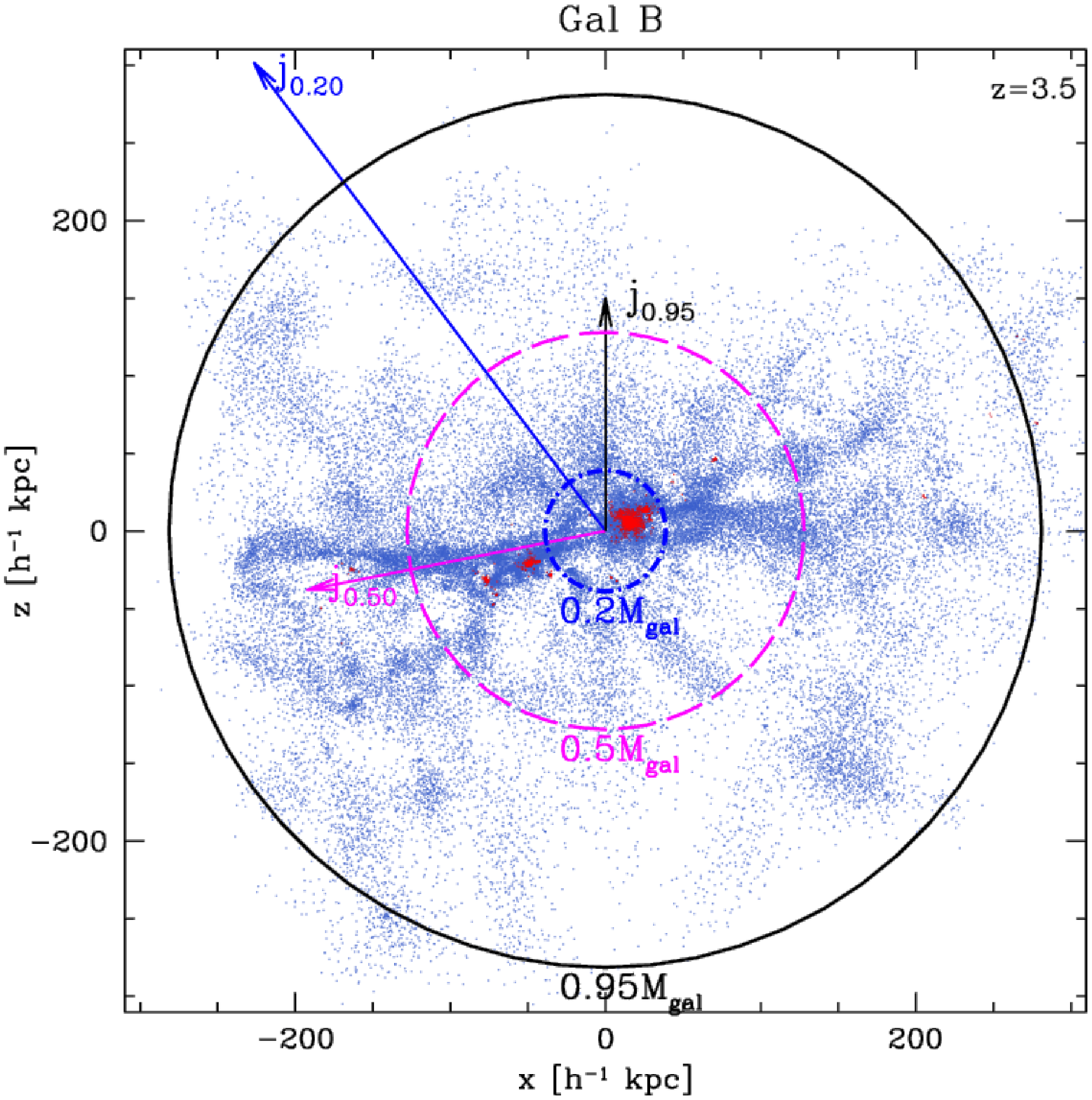}   
 \includegraphics[width=0.475\linewidth]{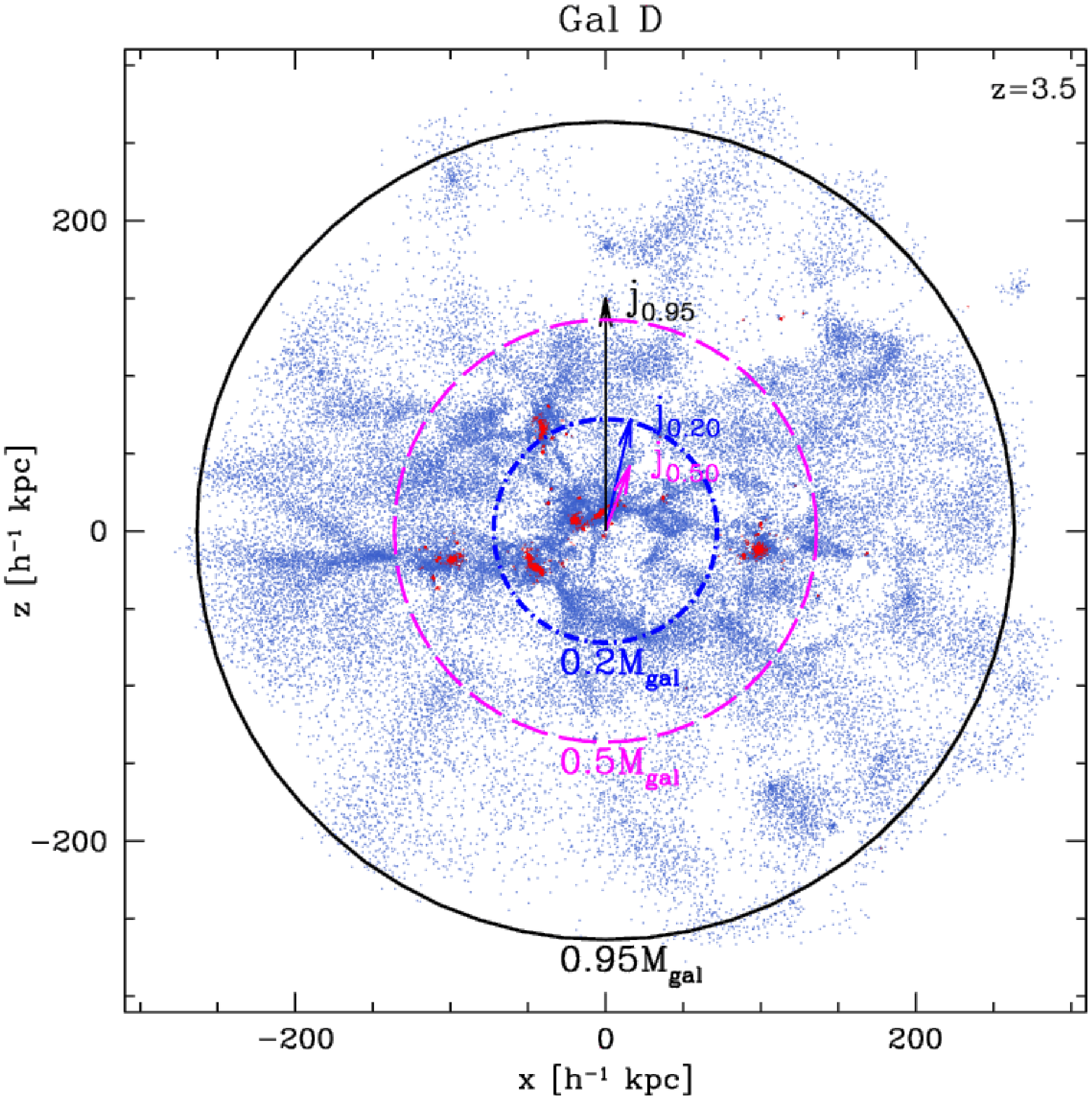}
 \caption{Projected particle distribution near turnaround time,
   $z=3.5$, of baryons that collapse to form, at $z=0$, galaxies B and
   D shown in Fig.~\ref{FigImage}. Stars already formed are shown in
   red, particles still in gaseous form in blue. Box sizes are in
   physical units. Concentric circles enclose $20\%$, $50\%$, and
   $95\%$ of the mass, and arrows indicate the angular momentum of all
   material enclosed within each radius. Arrow lengths are normalized
   to the total value, which defines the $z$ axis of the
   projection. Each panel is normalized separately, so that ${\bf j}_{0.95}$
   has equal length in both. Note the misalignment of the angular
   momentum of various parts of the system for the spheroid-dominated
   galaxy B. Angular momentum is more coherently acquired in the case
   of the disk-dominated galaxy D.}\label{fig:shells}
\end{figure*}
\end{center}

\begin{center} \begin{figure*}
 \includegraphics[width=0.475\linewidth]{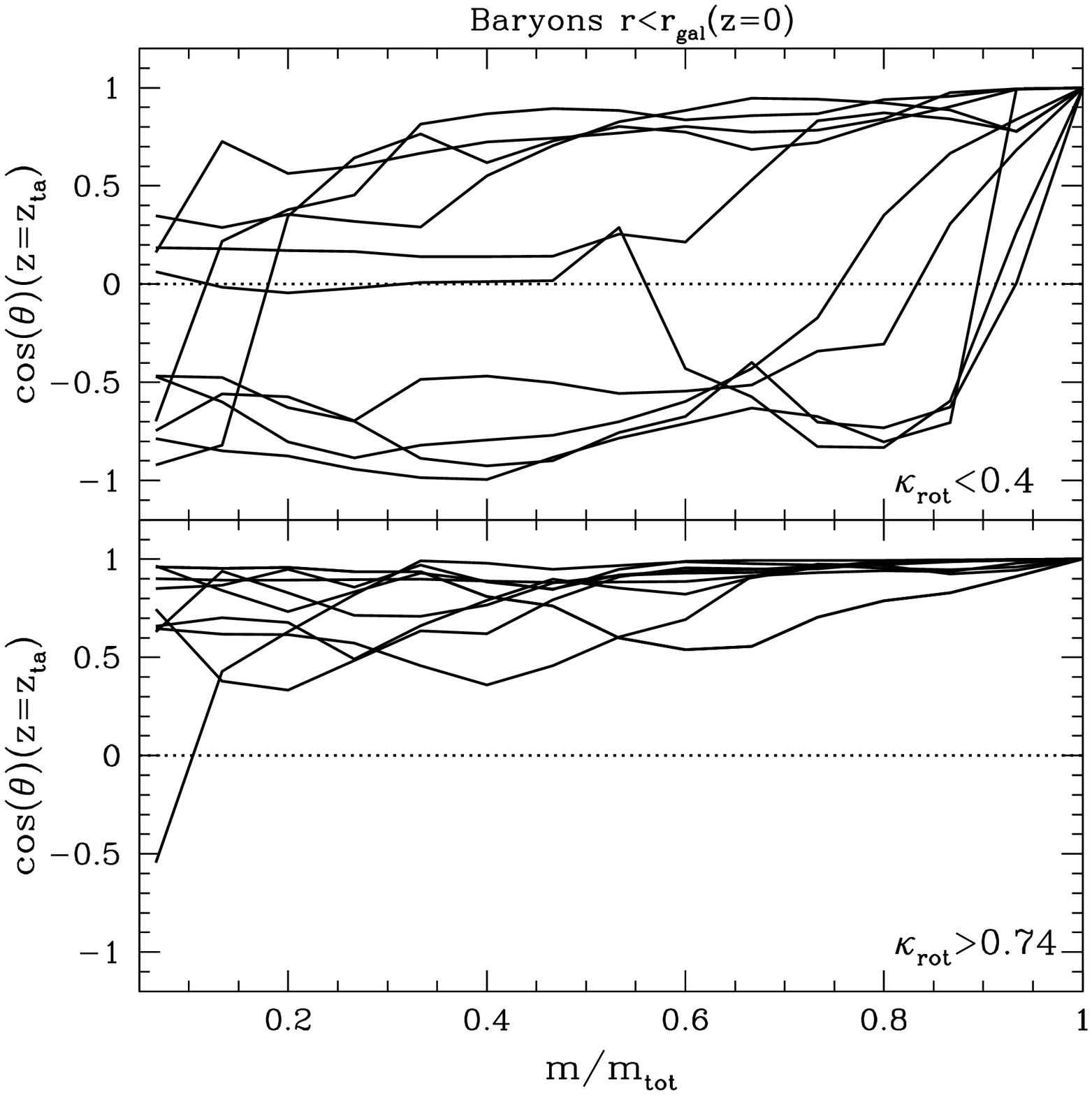} 
 \includegraphics[width=0.475\linewidth]{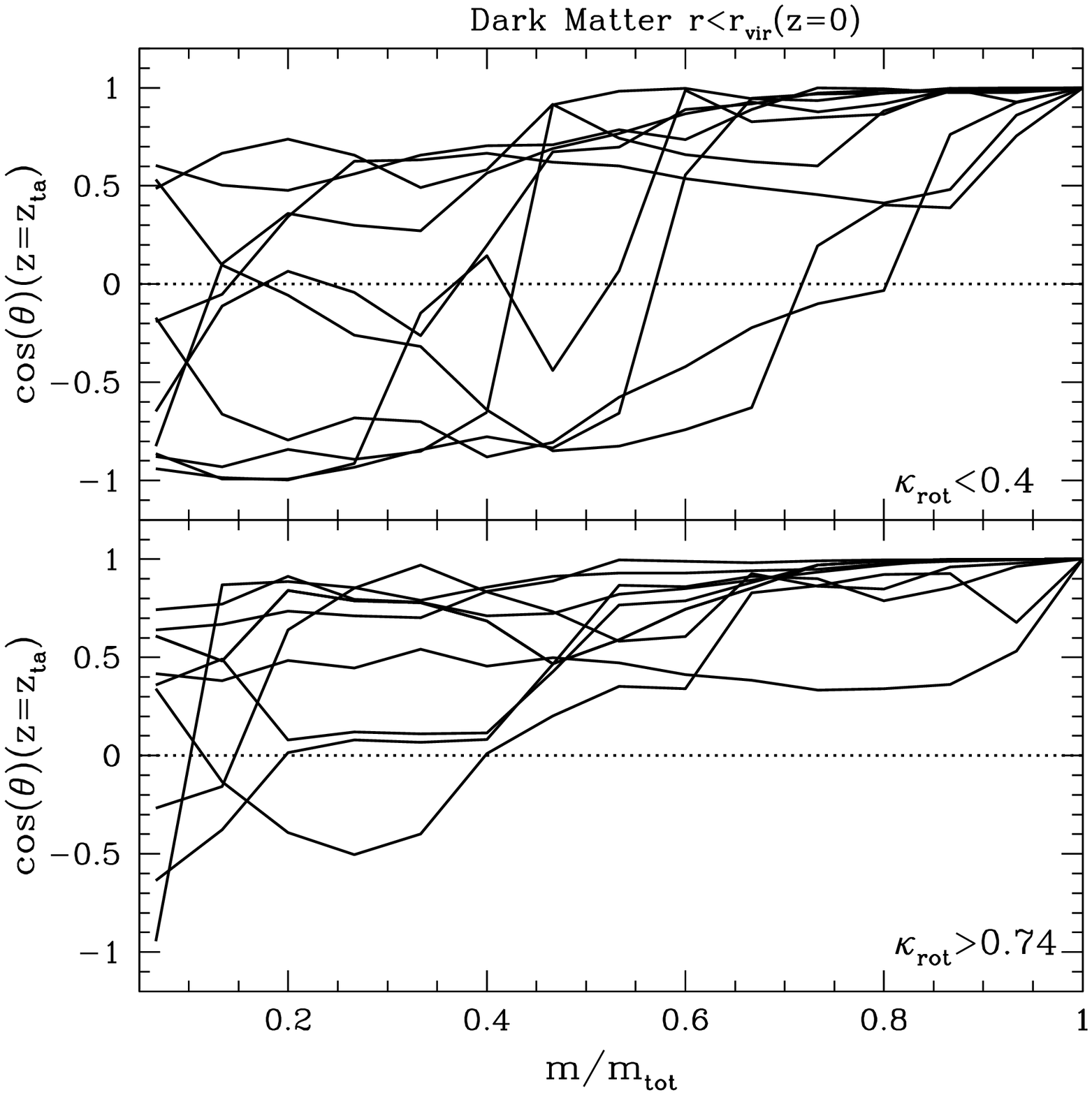} 
 \caption{Angle between the angular momentum enclosed within a given
   mass fraction, $m/m_{\rm tot}$, and the total spin of the system
   measured at the time of maximum expansion (turnaround) for twenty
   galaxies in our sample: the ten systems with highest and lowest
   values of $\kappa_{\rm rot}$ at $z=0$ and $f_{\rm acc}<0.1$,
   respectively.  The restriction in $f_{\rm acc}$ is included in
   order to focus on systems unaffected by merger events.  By
   definition, all curves approach unity as $m \rightarrow m_{\rm
     tot}$. Panels on the left correspond to all baryons within the
   galaxy radius, $r_{\rm gal}$, at $z=0$; those on the right to dark
   matter halo particles that are within $r_{200}$ at $z=0$. Note the
   strong misalignments between different parts of the system for
   galaxies that are spheroid-dominated at $z=0$, and the smooth
   alignment characteristic of the turnaround configuration of systems
   destined to become disks.}
\label{FigJzM}
\end{figure*}
\end{center}

\subsection{Dependence on galaxy history}
\label{SecHist}

Galaxy mergers can still in principle play a role in determining
morphology, if their importance is underestimated by the {\it halo}
merger parameter $\Delta M_{\rm lmm}$. Indeed, galaxies take longer to
merge than halos do, and, due to the large scatter in galaxy formation
efficiency, the mass ratio of galaxy mergers may differ substantially
from that of their surrounding halos.

We examine the importance of accretion on morphology more explicitly
in the middle-right panel of Fig.~\ref{fig:correlations}, where we plot
$\kappa_{\rm rot}$ vs $f_{\rm acc}$, the fraction of stars {\it
  accreted} by the galaxy; i.e., those formed in systems {\it other}
than the main progenitor of the galaxy. This is a direct measure of
the importance of accretion events in the build-up of the galaxy. Two
points are worth noting here: most galaxies form the majority
($>90\%$) of their stars {\it in-situ}, and there is no 
correlation between $\kappa_{\rm rot}$ and $f_{\rm acc}$. The accreted
fraction exceeds $25\%$ in only $5$ of our $100$ simulated galaxies;
overall, accretion events just seem to bring in too few stars to play a
significant role in the morphology of our simulated galaxies.

An interesting clue is provided by the thermodynamic history of the
gas before it is transformed into stars. This may be estimated simply
by tracking every stellar particle back in time and by recording the
maximum temperature, $T_{\rm max}$, reached before the particle
accretes into the galaxy and becomes eligible for star formation. If
$T_{\rm max}$ exceeds $10^{5.5}$ K, then in all likelihood it was
accreted by gradual cooling from a shock-heated, nearly hydrostatic
gas corona \citep[see, e.g.,][]{Crain2010,vandeVoort2011}.

The fraction of stars, $f_{\rm hot}$, whose parent gas particles went
through this phase correlates well with $\kappa_{\rm rot}$,
indicating that accretion of gas from the ``hot phase'' favours the
formation of disks (see bottom-right panel of
Fig. ~\ref{fig:correlations}). No disk-dominated galaxy (i.e. 
$\kappa_{\rm rot}>0.7$) forms unless $f_{\rm hot}$ exceeds
$40\%$. This is intriguing, since it runs counter recent proposals
that ``cold flows'', i.e., gas that gets accreted directly into the
galaxy without going through the hot phase, might promote the
formation of extended disks
\citep[e.g.,][]{Keres2005,Dekel2009,Brooks2009}. If anything, our
simulations suggest the opposite; the majority of stars in
spheroid-dominated galaxies originate in gas that accretes cold.

One reason for this result is illustrated in Fig.~\ref{figs:age_fhot},
which shows that there is a tight correlation between the fraction of
stars born from gas that cooled from a hot corona and the median
formation time of stars in a galaxy, $t^*_{50\%}$. Heating gas to a
hot corona before cooling delays its accretion and favors the late
assembly of a galaxy: the larger $f_{\rm hot}$ the later stars
form. Enhancing recent star formation promotes the formation of disks
and facilitates their survival until the present. This scenario,
although appealing, seems incomplete, given the presence of several
spheroid-dominated systems that form despite high $f_{\rm hot}$, late
$t^*_{50\%}$, and in the absence of significant merger activity.

\subsection{Dependence on spin alignment}
\label{SecAlign}

Our results so far suggest that the morphology of {\sc gimic} galaxies
is linked largely to internal mechanisms operating in individual galaxies
rather than to accretion-driven transformations. This has been
anticipated by semi-analytic models of galaxy formation, where secular
evolution driven by ``disk instabilities'', is thought to be an
important formation path for spheroids.  These instabilities are
assumed to be triggered when the self-gravity of a disk exceeds a
particular threshold. For fixed halo mass, as in our sample, this
should lead to noticeable correlations between the mass of the galaxy
and the importance of the spheroid. However, as we discussed above, no such
correlation is apparent.

Recalling the relation between $f_{\rm hot}$, $t^*_{50\%}$, and
$\kappa_{\rm rot}$, we look at the star formation history for further
clues. Fig.~\ref{fig:tform} plots the formation redshift, $z_{\rm
  form}$, of stars in the four galaxies shown in Fig.~\ref{FigImage}
versus the "circularity'' parameter, $\epsilon_j$, measured at $z=0$.
Points in black correspond to stars formed {\it in-situ} (i.e., within
the main progenitor) while those in red are accreted stars.

This figure shows that, as expected, star formation proceeds gradually
in disk-dominated systems. In spheroids, however, stars form in
separate episodes that leave behind stellar ``populations'' of
different ages and distinct angular momenta.  Since most stars form
{\it in-situ}, these populations in galaxies A and B are likely caused
by gas accretion events where the net angular momentum of one event is
misaligned with the others. On the other hand, disks tend to form out
of accreted gas that shares a common angular momentum direction. 

\begin{center} \begin{figure}
 \includegraphics[width=84mm]{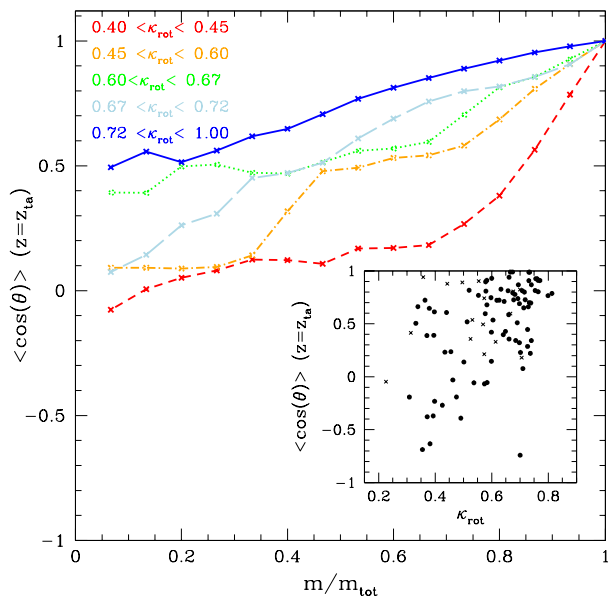} 
 \caption{Same as left-hand panel of Fig.~\ref{FigJzM}, but averaged
   over all galaxies grouped in bins of $\kappa_{\rm rot}$ (at $z=0$),
   as labeled. Note that the more disk-dominated a galaxy is at
   present, the more coherently aligned the spin is at the time of
   turnaround.  The inset panel shows $\langle \cos(\theta) \rangle$
   for individual systems, averaged over all mass shells at
   turnaround, as a function of $\kappa_{\rm rot}$. Galaxies where
   accretion events have played a minor role, i.e., $f_{\rm acc}<0.1$,
   are shown with solid circles, the rest with crosses.}
\label{FigJzMkrot}
\end{figure}
\end{center}

These considerations suggest that the final morphology of a galaxy is
imprinted early on, since the spin of the material
destined to form a galaxy is acquired at the time of maximum expansion
and changes little in the absence of merging \citep[see,
e.g.,][]{White1984,Navarro2004}. We illustrate this by studying the
angular momentum of galaxies B and D at $z=3.5$, which roughly
corresponds to the time of turnaround of both
systems. Fig.~\ref{fig:shells} shows the spatial distribution of all
baryons that will end up within $r_{\rm gal}$ at $z=0$.  Net angular
momentum is acquired through the interplay between the inertia tensor
of the mass distribution and the shear tensor due to the large-scale
distribution of surrounding matter, so that the direction of the
acquired spin usually aligns with the {\it intermediate axis} of the
mass distribution
\citep{Catelan1996,Porciani2002a,Porciani2002b,Navarro2004}. For
highly non-uniform spatial distributions, where the principal axes of
the inertia tensor can change direction abruptly, this effect may cause the net
angular momentum of different parts of the system to flip and
misalign.

We see from Fig.~\ref{fig:shells} that this is indeed the case for
galaxy B. Here the arrows indicate the direction and magnitude, at
$z=3.5$, of the specific angular momentum of the inner $20\%$, $50\%$
and $95\%$ of the baryons that end up in the galaxy at $z=0$. The
length of the arrows is normalized to the total, which is chosen by
construction to coincide with the $z$ axis of the projection. The
angular momenta of different parts of the system are clearly
misaligned: the angle between ${\bf j}_{0.2}$ and ${\bf j}_{0.5}$ is
$85$ degrees, and that between ${\bf j}_{0.5}$ and ${\bf j}_{0.95}$ is
$\sim 100$ degrees. Since gas further out in Fig.~\ref{fig:shells}
takes longer to accrete,
newly assembled material will be misaligned with the rest, leading to
the formation of distinct populations of stars (shown in
Fig.~\ref{fig:tform}) that will tend to destabilize any existing disk
and to cancel out the net angular momentum of the system; leaving in
place a slowly-rotating stellar spheroid.

On the other hand, the spins of different parts of the system are very
well aligned in the case of the disk-dominated galaxy D, as shown in
Fig.~\ref{fig:shells}. This coherence in the angular momentum at
turnaround allows newly accreted material to settle into a stable disk
where star formation can proceed gradually and smoothly.

We show in Fig.~\ref{FigJzM} that this result applies to the majority
of spheroid- and disk-dominated galaxies in our sample. Here we plot,
for the ten galaxies with highest and lowest values of $\kappa_{\rm
  rot}$ where accretion has played a minor role ($f_{\rm acc}<0.1$),
the cosine of the angle $\theta$ between the angular momentum of a
given enclosed mass fraction $m/m_{\rm tot}$ and the direction of the
total spin of the system. By construction, each curve in
Fig.~\ref{FigJzM} is constrained to approach unity as the enclosed
mass approaches $m_{\rm tot}$.

It is clear from this figure that different regions of systems
destined to form spheroid-dominated galaxies have, at turnaround,
large misalignments in their acquired spin. Indeed, in many cases the
inner regions counterrotate (i.e., $\rm cos(\theta)<0$) relative to
the outer regions of the system. This is not the case for systems that
become disk-dominated which, in general, show coherence in the
alignment of the spin axis. In these cases, the enclosed specific
angular momentum increases roughly linearly with enclosed mass
fraction. As Fig.~\ref{FigJzMkrot} shows, the same result applies to
all galaxies in our sample: despite the large scatter, on average, the
degree of alignment at turnaround increases gradually with the
importance of the disk component in the morphology of a galaxy at
$z=0$.

The right panels in Fig.~\ref{FigJzM} show that a similar assessment
applies to the dark matter halo surrounding these galaxies. The halos
of spheroids also show, at turnaround, stronger misalignments than the
halos that host disk galaxies at $z=0$.  This is encouraging, since it
implies that it might be possible to use the angular momentum
properties of a dark matter halo at turnaround to ``predict'' the
morphology of its central galaxy at $z=0$. We emphasize, however, that
the trends we highlight here, although well defined, are relatively
weak, so the correspondence between early halo properties and final
galaxy morphology is likely to apply statistically rather than to
individual systems.

Quantitatively, the dependence of present day morphology on spin
alignment at turnaround is shown in the inset panel of
Fig.~\ref{FigJzMkrot}. Here we show, for each individual system,
$\kappa_{\rm rot}$ versus the average cosine of the angle, at
turnaround, between the angular momentum enclosed by different mass
shells and the total. The trend is clear: systems with better-aligned
spins at turnaround tend to be more disk-dominated at present. The
trend is even stronger when considering only systems where mergers
have played a minor role ($f_{\rm acc}<0.1$, solid points).  The
correlation coefficient is $r_s=0.49$ (with significance $\Delta \sim
2 \times 10^{-6}$) and confirms our earlier conclusion that
significant misalignment in the distribution of angular momentum can
lead to the formation of spheroid-dominated systems in the absence of
significant merger events.

\begin{center} \begin{figure}
 \includegraphics[width=84mm]{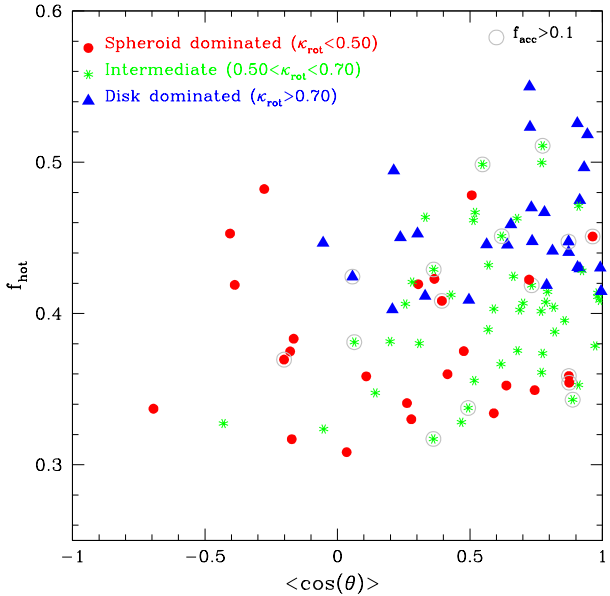} 
 \caption{Fraction of stars born from gas accreted from the ``hot
   phase'' as a function of the alignment parameter
   $\langle\cos(\theta)\rangle$ at the time of
   turnaround. Color-coding is the same as in
   Fig.~\ref{figs:age_fhot}. Note that, in general, disk-dominated,
   spheroid-dominated, and intermediate systems occupy different
   regions in this diagram. See text for further discussion.}
\label{figs:fhot_align}
\end{figure}
\end{center}

\subsection{Spin alignment vs mode of accretion}

The results of the above subsections suggest that there are (at least)
two mechanisms responsible for the morphology of {\sc gimic} galaxies:
the alignment of the spin acquired by various parts of a galaxy and
the mode of gas accretion.  Fig.~\ref{figs:fhot_align} shows that
these two effects are approximately independent of each other: there
is no obvious correlation between the fraction of gas accreted from
the hot phase and the alignment parameter
$\langle\cos(\theta)\rangle$.  However, disk-dominated systems (blue
triangles) separate from spheroid-dominated ones (red circles) fairly
neatly in this plane, suggesting that $\kappa_{\rm rot}$ is determined
by a combination of the two mechanisms.

Large $f_{\rm hot}$ favors disk formation: gas shock-heated into a
nearly-hydrostatic corona of hot gas is forced to homogenize its
rotational properties before accretion, providing the forming galaxy
with a gradual supply of gas that shares the same spin axis
\citep[see, e.g.,][]{Brook2012}. Disk-dominated systems then form when
$f_{\rm hot}$ is large, unless the spin misalignments are large enough
to disrupt them: no disk-dominated system forms when
$\langle\cos(\theta)\rangle<0$, even for relatively large values of
$f_{\rm hot}$.

Spheroids, on the other hand, form primarily when cold gas accretion
prevails ($f_{\rm hot}<0.4$): gas that flows along
distinct filaments cannot interact hydrodynamically before accretion
and will often have misaligned net spins. Each accretion
event then results in the formation of a ``population'' of misaligned stars
that will tend to destabilize any existing disk and to cancel out the
net angular momentum of the system, leaving in place a slowly-rotating
stellar spheroid.  Intermediate systems, in general, result when the deleterious
effects on stellar disks of cold accretion events are mitigated by
well-aligned spins.

In other words, spheroid-dominated galaxies in our sample do not
originate from disk instabilities triggered by self-gravity, as
envisioned by semi-analytic models, but rather by the accretion of gas
that settles on off-axis orbits relative to earlier accreted
material. This has been seen in earlier work
\citep[e.g.,][]{Brook2008,Scannapieco2009}, and might be related to
sudden changes in the orientation of the dark matter halos as
discussed in \citet{Bett2011}. Its relevance to the formation of the
whole class of spheroidal galaxies, in the absence of merging, has not
yet been recognized and emphasized.

We hasten to add that the importance of this mechanism for the
formation of spheroids might depend on halo mass, and that we explore
only a narrow range here: $0.5<M_{\rm 200}/10^{12} \, h^{-1}
M_\odot<1.5$. This caveat might be particularly relevant in the case
of the most massive spheroids, where merging likely plays a more
important role \citep[e.g. ][]{Parry2009, DeLucia2011,
  Feldmann2010}. Caution must also be exercised when extrapolating the
link between morphology and the fraction of stars born from hot
accretion. Disk-dominated galaxies might still form out of cold
accretion if, for example, the most recent episode of accretion
supplies most of the mass of the system. Repeated cold accretion
events may hinder disk formation, but a single major event may very
well facilitate it.

\section{Summary}
\label{sec:concl}

We use gasdynamical cosmological simulations of galaxy formation to
study the origin of different galaxy morphologies in the $\Lambda$CDM
cosmogony at redshift $z=0$.  The {\sc gimic} simulation series covers
a large volume and has a resolution high enough to study the structure
and kinematics of the stellar components of $100$ central galaxies in
Milky Way-sized halos. We focus our analysis on the origin of galaxy
morphology, somewhat narrowly defined as the relative importance of
rotational support vs velocity-dispersion support (the
disk-to-spheroid ratio) in the structure of the galaxy. Our main
results may be summarized as follows:

\begin{itemize}

\item The simulated galaxies span a wide range of morphological types,
  from rotation-free spheroids to almost pure disk galaxies where
  fewer than $5\%$ of all stars are in counterrotating orbits. Disks
  have roughly exponential stellar surface density profiles and flat
  rotation curves, whereas spheroids are dense stellar systems that
  can be approximated by de Vaucouleurs' $R^{1/4}$ profiles. The resemblance
  with real galaxies suggests that it should be possible to gain
  insight into the origin of galaxy morphology by studying the
  mechanisms responsible for the relative importance of disks and
  spheroids in {\sc gimic} galaxies.

\item The morphology of simulated galaxies seems mostly unrelated to
  the spin or assembly history of their surrounding dark matter
  halos. Most stars form {\it in-situ} and comprise on average about
  $\sim 40\%$ of all available baryons in the halo. Most baryons in a
  halo therefore end up not making part of the central galaxy, which
  helps to explain the weak correlation between the properties of
  halos and those of central galaxies.  Contrary to simple
  expectations, disks form in halos with low and high angular momenta,
  and spheroids form even in galaxies where most stars form {\it
    in-situ}, suggesting a formation path for spheroids that does not
  rely on merging.

\item The star formation history provides an interesting clue to the
  origin of morphology. Disks tend to have young stars, and to form
  gradually over long periods of time. This is because gradual cooling
  from a hot corona delays the accretion of gas and promotes late star
  formation. 

\item Star formation in spheroids proceeds episodically, leaving
  behind populations of stars of similar age but distinct
  kinematics. These populations originate from the accretion of gas
  whose angular momentum is misaligned relative to that of
  earlier-accreted material. The misalignment destabilizes any
  pre-existing disk, prompts the rapid transformation of gas into
  stars, and reduces the net rotational support of the system.

\item Since angular momentum is largely acquired at the time of
  maximum expansion of the material destined to form a galaxy,
  a good indicator of morphology at $z=0$ is the coherence in the
  alignment of the net spin of various parts of the system at the time
  of turnaround. Spheroid-dominated galaxies form in systems where
  misalignments are substantial whereas disks form in systems where the
  angular momentum of all mass shells is roughly aligned. 

\item The final morphology of a galaxy results from the combined
    effects of spin alignment and of hot/cold gas accretion.
    Disk-dominated objects are made of stars formed predominantly {\it
      in situ}, and avoid systems where most baryons were accreted
    cold, or those where spin misalignments are extreme. On the other
    hand, direct filamentary accretion of cold gas, especially when
    accompanied by substantial spin misalignments, favours the
    formation of slowly-rotating spheroids, which may thus form even in
    the absence of mergers.  

\end{itemize}

Our results suggest a new scenario for the origin of $\sim L_*$
stellar spheroids that does not rely on merging. This scenario, once
developed more thoroughly, should offer a number of predictions
falsifiable by observation. For example, the episodic nature of star
formation in spheroids envisioned here is expected to leave behind
overlapping populations of stars of distinct age, kinematics, and,
possibly, metallicity) that survive to the present because of the
paucity of mergers. We plan to explore the observational signatures of
these populations in future work.

The scenario we propose here also offers clues to the origin of pure
disk galaxies. A number of our simulated galaxies have virtually no
``classical'' spheroid, with fewer than $5\%$ of their stars in
counterrotating orbits. These galaxies form either in systems where
spin alignment is extraordinarily coherent or where most of the
baryons in the galaxy get accreted late from a hot corona. Although at
this time limited numerical resolution precludes a more detailed
study, we plan to use these clues to resimulate some of these systems
at higher resolution with the goal of shedding light into the origin
of bulgeless galaxies.

We emphasize that, although coherent spin alignment at early times is
clearly an important clue, it should be considered as one ingredient
of the complex process that determines the morphology of a galaxy.
Strong feedback, for example, may expel baryons from galaxies and
cycle them through a hot corona before they get re-accreted and turned
into stars, potentially erasing the spin alignment dependence at
turnaround we report here. Furthermore, aligned spins in the accreting
gas might not be enough to ensure the survival of a stellar disk,
especially if the dark matter halo is strongly triaxial and its
principal axes are not coincident with the disk. Finally, although
mergers are rare in the mass range we explore here, they likely play a
more important role in the formation of more massive spheroids. Until
simulations can reproduce not only the properties of individual
systems, but the full statistical distribution of galaxy morphologies
and their dependence on mass and environment, it is likely that a full
understanding of the origin of galaxy morphology will remain beyond
reach.

\section*{Acknowledgements}
\label{acknowledgements}

We thank the anonymous referee for a prompt and useful report.
We thank Ian McCarthy for his contribution to the simulations used in
this work, which were run at the Darwin HPC of Cambridge
University. LVS is grateful for financial support from the
CosmoComp/Marie Curie network.  CSF acknowledges a Royal Society
Wolfson research merit award and ERC Advanced Investigator grant
COSMIWAY. This work was supported in part by an STFC rolling grant to
the ICC.

\begin{center}
\begin{table} 
  \caption{Summary of main properties for Gal A-D in Fig.~\ref{FigImage} and \ref{FigVcSigmaProf}.
    Rows correspond to the virial mass $M_{200}$; galactic mass in stars $M_{\rm str}$; 
    gas $M_{\rm gas}$ and gas fraction $f_{\rm gas}$; peak circular velocity 
    $V_{\rm max}$; the circular velocity
    measured at the galactic radius $V_c(r=r_{\rm gal})$; the degree 
    of rotational support $\kappa_{\rm rot}$; and 
    the fraction of the stars in counter-rotating orbits $f_c$. }
\begin{tabular}[width=0.85\linewidth,clip]{|l|c|c|c|c|}
\hline
 Property & Gal A & Gal B & Gal C & Gal D \\
\hline
$M_{200} [10^{12} h^{-1}M_\odot]$ & 1.18 & 0.98 & 0.77 & 1.04 \\ 
$M_{\rm str} (r<r_{\rm gal}) [10^{10} h^{-1}M_\odot]$ & 8.98 & 7.25 & 5.42 & 6.31 \\
$M_{\rm gas} (r<r_{\rm gal}) [10^{10} h^{-1}M_\odot]$ & 0.51 & 0.72 & 1.00 & 1.86 \\ 
$f_{\rm gas} (M_{\rm gas}/(M_{\rm gas}+M_{\rm str}))$ & 0.05 & 0.09 & 0.15 & 0.23 \\
$V_{\rm max}$ [km/s] & 471 & 383 & 300 & 280 \\
$V_c(r=r_{\rm gal})$ [km/s] & 187 & 172 & 160 & 178 \\
$\kappa_{\rm rot}$ & 0.31 & 0.47 & 0.61 & 0.76 \\
$f_c$ & 0.48 & 0.33 & 0.12 & 0.06 \\
\hline
\hline
\end{tabular}
\label{tab:table1}
\end{table}
\end{center}

\begin{center}
\begin{table} 
  \caption{Spearman rank correlation coefficients, $r_s$, between $\kappa_{\rm rot}$ and
    the halo/galaxy properties shown in Figs.~\ref{fig:correlations}
    and \ref{FigJzMkrot}. The second column
    shows the two-sided significance of its deviation from zero, as computed by the IDL
    subroutine {\tt r-correlate}: smaller values 
    indicate more significant correlations.}
\begin{tabular}[width=0.85\linewidth,clip]{|l|r|r|}
\hline
 Property & $r_s$ & $\Delta$ \\
\hline
$t_{50\%}$ & $0.05$ & $0.60$ \\
$\Delta_{\rm lmm}$ & $-0.06$ & $0.57$ \\
$\lambda'$ & $0.29$ & $2.8\times 10^{-3}$ \\
$\eta_{gal,*}$ & $-0.30$ & $1.9 \times 10^{-3}$ \\
$f_{\rm acc}$ & $0.06$ & $0.55$ \\
$f_{\rm hot}$ & $0.55$ & $1.6\times 10^{-9}$ \\
$<\rm cos(\theta)>$ & $0.41$ & $2.0 \times 10^{-5}$\\
\hline
\hline
\end{tabular}
\label{tab:correlation}
\end{table}
\end{center}

\bibliography{master}

\end{document}